\documentclass[journal]{IEEEtran}
\usepackage{amsmath,amssymb,amsfonts}
\usepackage{algorithmic}
\usepackage{algorithm}
\usepackage{array}
\usepackage[caption=false,font=normalsize,labelfont=sf,textfont=sf]{subfig}
\usepackage{textcomp}
\usepackage{stfloats}
\usepackage{url}
\usepackage{verbatim}
\usepackage{graphicx}
\usepackage{cite}
\usepackage{comment}
\usepackage{booktabs}

\usepackage{siunitx}
\usepackage{breqn}
\usepackage{hyperref}
\usepackage{svg}
\usepackage{float}

\begin{document}

\title{Vital Signs Monitoring with mmWave OFDM JCAS System}
\author{Jakub Dobosz, Maximilian Engelhardt, Diego Dupleich, Maciej Stapor, and Pawel Kulakowski
\thanks{J. Dobosz is with the Institute of Telecommunications, AGH University of Krakow, Poland (e-mail: jakub.dobosz01@gmail.com).}
\thanks{M. Engelhardt is with the Technische Universität Ilmenau (e-mail: maximilian.engelhardt@iis.fraunhofer.de)}
\thanks{D. Dupleich is with the Technische Universität Ilmenau (e-mail: diego.dupleich@tu-ilmenau.de)}
\thanks{M. Stapor is with Jagiellonian University Medical College, Faculty of Medicine, Institute of Cardiology, Department of Interventional Cardiology, St. John Paul II Hospital, Krakow, Poland (e-mail: maciej.stapor@uj.edu.pl).}
\thanks{P. Kulakowski (corresponding author) is with the Institute of Telecommunications, AGH University of Krakow, Poland (e-mail: kulakowski@agh.edu.pl).}}

%\markboth{Submitted to IEEE Access}%
%{J. Dobosz \MakeLowercase{\textit{et al.}}: Vital Signs Monitoring with mmWave OFDM JCAS System}


\maketitle

\begin{abstract}
Wireless techniques for monitoring human vital signs, such as heart and breathing rates, offer a promising solution in the context of joint communication and sensing (JCAS) with applications in medicine, sports, safety, security, and even the military. This paper reports experimental results obtained at the Fraunhofer Institute for Integrated Circuits in Ilmenau, demonstrating the effectiveness of an indoor orthogonal frequency-division multiplexing (OFDM) JCAS system for detecting human heart and breathing rates. The system operated in a bistatic configuration at an FR2 frequency of 26.5 GHz with a variable bandwidth of up to 1 GHz. Measurements were taken under various scenarios, including a subject lying down, sitting, or walking, in both line-of-sight and non-line-of-sight conditions, and with one or two subjects present simultaneously. The results indicate that while vital sign detection is generally feasible, its effectiveness is influenced by several factors, such as the subject’s clothing, activity, as well as the distance and angle relative to the sensing system. In addition, no significant influence of bandwidth was detected since the vital signs information is encoded in the phase of the signal. 
\end{abstract}


\section{Introduction}
\IEEEPARstart{T}{he} continuous and contactless monitoring of vital signs, such as heart rate (HR) and breathing rate (BR), has recently gained attention due to its massive potential primarily in healthcare and elderly care, but also in security, safety, and personal applications. This technology is valuable for long-term vital sign monitoring, sleep tracking, and as an assistance tool for seniors. It could also be used as a tool for detecting driver fatigue without the privacy concerns associated with camera-based solutions. Furthermore, it has a large variety of security applications related to the detection of intruders, or even recognizing people via their breathing and heartbeat patterns. A potential advantage of the technology over other sensors is its independence from the lighting conditions, which could be an issue when operating at night. 

The currently used methods require direct contact with the examined person. It applies to both professional systems used in hospitals, such as electrocardiography (ECG), pulse oximeters, and capnography systems, as well as consumer wearable devices, including smartwatches and heart-monitoring belts. Some of the contact-based sensors can be particularly inconvenient for individuals with allergies or sensitive skin. For example, they may not be suitable for infants and elderly patients. The reliable contactless monitoring of vital signs would be much more comfortable, as patients could be monitored constantly without the need for physical interaction with the device. Radio transceivers operate by emitting electromagnetic waves that can easily penetrate thin, non-conductive materials such as clothing or blankets. Therefore, such systems can monitor patients and people in real-world conditions without degrading the accuracy or reliability of the measurements. 

Furthermore, the potential of sensing is explored in the context of emerging advancements in sixth-generation (6G) mobile communications and next-generation Wi-Fi technologies. In these systems, joint communication and sensing (JCAS) is envisioned to seamlessly merge communication and sensing functionalities, thereby transforming wireless networks into multi-purpose platforms. This integration is expected to unlock a wide range of innovative applications, particularly in the fields of healthcare and security, paving the way for smarter, more responsive, and context-aware environments.

\subsection{Principles of Contactless Respiration Rate and Heart Rate Monitoring}
\label{sec:principles}

Contactless monitoring of vital signs relies on detecting periodic movements of the body caused by breathing and heart activity. Radio-frequency signals are transmitted towards the person, and a part of them reflects in the direction of the receiving antenna. The periodic movements of the human body modulate the phase of the received signal. The path length decreases during inhalation as the chest moves closer to the device and increases during exhalation. This, in turn, causes a phase shift in the received signal. 

The detected signal is, for the most part, reflected from the skin of the person and not from the internal organs. It results from the high attenuation of EM waves by body tissues. Additionally, for radio waves below 18~GHz, the magnitude of the reflection coefficient $\Gamma_{\text{air/skin}}$ between air and skin is above 0.7, which means that the air-skin interface reflects more than 70\% of the energy~\cite{Aardal2013}. In the case of millimeter waves (mmWaves), less energy is reflected, but these waves are predominantly absorbed in the first few centimeters of the tissue. Therefore, even though the movements of the heart are greater than those of the skin, the amount of signal reflected from the heart walls is so low that it has little impact on the received signal~\cite{Aardal2013}. 

For a typical adult, the chest moves about 1 to 12 mm due to breathing and 0.01 to 0.5 mm due to heartbeats. The phase change ($\Delta \phi$) is given by:

\begin{equation}
    \Delta \phi = \frac{4 \pi \Delta d}{\lambda}
\end{equation}

where $\Delta d$ is the chest displacement, and $\lambda$ is the wavelength. Equation (1) shows that the shorter the wavelength, the greater the phase change for a given displacement. Therefore, mmWave systems, which operate at shorter wavelengths, can measure phase differences more precisely than when operating at lower frequencies, thus enabling the detection of movements of the human body.

Apart from the measurement based on phase variation, an alternative approach for vital signs monitoring involves tracking changes in the magnitude of the received signal caused by vital signs. However, magnitude-based measurements are generally considered less robust because the small movements involved cause only minor variations in signal power, which are difficult to measure \cite{Alizadeh2019}. Measuring HR using signal strength change can be practically challenging, but it is more feasible for applications such as tracking the number of people in a room.  

\subsection{Paper Scope and Contribution}
\label{sec:contribution}

In this paper, we present the results of experiments on indoor measurement of human VS with a bistatic mmWave OFDM radar. Our study contributes to the field of wireless OFDM-based VS monitoring by providing a comprehensive evaluation of the  performance in a wide range of realistic scenarios, identifying the challenges associated with the technology and its operational limits. Our approach leverages OFDM signals, the waveforms commonly employed in modern communication systems, and a bistatic set-up to evaluate the effectiveness of vital sign monitoring within the context of JCAS. 

%Additionally, our approach utilizes the raw OFDM waveform, which is less common compared to using Wi-Fi CSI information from selected subcarriers. Our approach offers the ability to extract more accurate measurements because of the extended instantaneous bandwidth resulting from the use of all the subcarriers. 

In the next section, we provide a medical perspective on the clinical importance of contactless VS monitoring of patients, and we also discuss security applications. This section is followed by an overview of related work in the field of human VS measurement with the propagation of EM waves. Then a description of the measurement setup and scenarios used in our experiments is given, followed by a discussion of the utilized signal processing methods and obtained results. The paper is concluded with a section with conclusions and suggestions on possible directions for future research.

\section{Applications of Contactless Vital Signs Monitoring}
\label{sec:applications}

Wireless monitoring of vital signs is transforming patient care across various healthcare environments, including cardiology, internal medicine, surgical wards, nursing homes, long-term care facilities, and home settings for seniors. This technology enables continuous, real-time tracking of physiological parameters, facilitating the early detection of patient deterioration and the timely implementation of interventions. It has also numerous security applications, including intruder detection and people identification in restricted areas. However, VS monitoring also raises privacy concerns related to people’s awareness of being wirelessly identified and tracked. 

\subsection{Hospital and Post-ICU Care}

In hospital settings, studies have demonstrated that continuous vital signs monitoring using wearable wireless devices can lead to improved patient outcomes. Wireless monitoring has been associated with shorter hospital stays, fewer nurse-to-physician calls in surgical patients, and a reduction in rapid response team calls, suggesting improved early recognition and management of clinical decline \cite{Leenen2024}.

In cardiology, a particularly relevant scenario involves patients who have been recently discharged from the intensive care unit (ICU) following an acute cardiac event. Although these patients are deemed stable enough to leave critical care, they often remain at high risk for early deterioration due to arrhythmias, volume shifts, or hemodynamic instability.

Wireless vital signs monitoring provides a valuable safety net during this vulnerable post-ICU transition phase. Continuous tracking of key parameters can detect subtle changes that precede clinical deterioration. Several studies support this approach, showing that continuous monitoring in step-down units can significantly reduce `failure-to-rescue' events and unplanned ICU readmissions \cite{Leenen2024}.

Moreover, the use of mobile or wearable wireless systems facilitates early discharge planning. It enables step-down care units or even home-based telemonitoring programs to safely monitor cardiac patients while reducing hospital stay durations and overall costs. This strategy aligns with modern heart failure and post-acute care guidelines that emphasize early mobilization and structured follow-up \cite{Clark2023}.

\subsection{Long-Term Care and Home Monitoring}

In long-term care facilities and home settings, wireless monitoring systems offer a way to maintain continuous oversight of elderly patients’ health, reducing the need for frequent in-person visits and enabling prompt response to early warning signs. A recent study highlighted the potential of wearable wireless sensors in out-of-hospital settings, noting that such devices can signal when a person should seek additional care, thereby improving outcomes for seniors living independently. Personalized health-monitoring systems that combine rule-based logic with case-based reasoning may also enhance tailored care and early intervention \cite{Ahmed2015}.

Devices for continuous vital sign monitoring may benefit various groups. In hospital inpatients, this allows for accurate and remote monitoring while minimizing disruptions and reducing the need for manual measurements. For infectious disease patients, they help limit exposure risk for healthcare workers. Furthermore, they increase patient mobility and comfort by reducing the number of wires and sensors needed \cite{Weller2018}.

Depending on the setting and facility, vital signs are currently monitored either manually or via wired monitors. Wireless devices such as wristbands, armbands, patches, and Holter monitors provide more flexibility and continuity of care. These benefits are especially valuable in both hospital and home care environments, protecting healthcare providers and family caregivers alike.

\subsection{Future Directions in Healthcare}

A promising future direction for wireless vital signs monitoring is the development of systems that do not require the patient to wear any device. Removing the need for physical sensors would enhance comfort and ease of use, particularly for the elderly or individuals with cognitive impairments. It would also address common challenges such as forgetting to wear the device, accidental dislodgement, or the need for regular charging and maintenance. This contactless approach could simplify monitoring workflows, reduce caregiver burden, and provide a more seamless and less intrusive experience for patients—particularly in long-term care and home settings \cite{Clark2023, Ahmed2015}. With the rapid technological improvement and growing demand for convenient alternatives to existing contact-based health monitoring systems, the U.S. market for vital signs monitoring devices is projected to exceed USD 11 billion by 2028 \cite{ResearchMarkets2023}.

\subsection{Safety and Security Applications}

Radio-frequency (RF) signals generated by communication systems can also be leveraged for the detection of intrusions within secured areas, eliminating the need for dedicated sensing devices such as cameras or specialized sensors, which may introduce additional infrastructure costs or raise privacy concerns, \cite{10107610}. Following the detection of an anomalous presence, subsequent signal analysis can facilitate the classification of the source as human, animal, or inanimate object, among others. In this context, physiological parameters such as heart rate and respiratory rate, which can be inferred from RF signal variations, provide critical discriminatory features. Moreover, these physiological indicators can be further utilized to estimate affective states, such as agitation or stress, thereby enabling an assessment of the intentions or potential threat level posed by an intruder.

In occupational settings, as logistic centers or industrial facilities, similar RF-based monitoring techniques present significant advantages. For example, the identification of personnel in restricted areas, as well as the non-intrusive assessment of worker stress levels, can be achieved without the deployment of additional sensing infrastructure, relying solely on the analysis of pre-existing communication signals.

\subsection{Privacy Aspects}

While the technology of contactless VS monitoring offers numerous benefits and a wide range of applications, it also raises significant privacy concerns. It can be easily used to monitor people without their explicit permission, or even keeping them unaware of the monitoring process. It is not far from a scenario when an employer monitors the stress level of his employees without their consent. Furthermore, breathing and heartbeat patterns of people can be recorded and they can be tracked and recognized among others, again without their knowledge. All these potential implications should be carefully considered as this technology matures and specific commercial solutions become available.

\section{Related Work}
\label{sec:related}

One of the most promising technologies for contactless measurement of human VS is the frequency-modulated continuous wave (FMCW) radar. FMCW radars are commercially available and widely utilized, especially in the automotive industry, as part of Advanced Driver-Assistance Systems (ADAS). The non-linear components of the system, namely power amplifiers and low-noise amplifiers, operate in their linear region in FMCW radars. This is a result of the fact that the FMCW signal has a constant envelope with a peak-to-average power ratio (PAPR) of 0 dB~\cite{alizadeh2019remote}. In the literature, different frequencies have been tested, including 9.6 GHz, 24 GHz, and 77 GHz, with the 77–81 GHz band being the most widely used for this purpose. FMCW radars can estimate the distance to a target with the range resolution of 
$\Delta R = c / 2 B$, where $c$ is the speed of light in vacuum, and $B$ is the bandwidth in Hz. This imposes high bandwidth requirements, especially in the context of VS monitoring applications, where body movement resulting from the heartbeat and breathing is on the order of centimeters or even millimeters. 

The article~\cite{alizadeh2019remote} presents the results of experiments conducted using a Texas Instruments mm-wave FMCW radar (AWR1443) operating in the 77–81 GHz frequency band. The experiment utilizes a single transmitter-receiver (Tx/Rx) pair with a chirp duration of 64~µs and a repetition period of $T_c = 50$~ms. The slope of each chirp is 70~MHz/µs with a bandwidth of 3.99~GHz. The maximum unambiguous distance between the patient and the radar is 2.14~m, which means that for greater distances, the radar will report values modulo 2.14~m. The measurements of BR and HR are taken after gathering samples from 256 chirp sequences, which gives the observation duration of 12.8~s. For optimization, the observation window utilizes 128 samples from the previous observation, resulting in new BR and HR measurements being effectively produced every 6.4 seconds. The study examines the performance of the radar in a scenario where the patient lies on their bed, facing up at the radar, which is mounted above. A wearable device is used as a reference for the tested device. The FMCW radar achieved 80\% accuracy for HR estimation and 94\% accuracy for breathing rate at a distance of 1.7~m with an output power of 15.8~mW. The results are comparable with those of others, as article~\cite{wang2015novel} reports achieving accuracies of 87.2\% and 91.08\% in the estimation of HR and BR, respectively. The setup included ultra-wideband 80 GHz FMCW with 10 GHz of bandwidth. 

An interesting approach for differentiating between multiple closely spaced subjects is presented in ~\cite{lee2019novel}. The available bandwidth of the 24~GHz FMCW radar used in the study was 250~MHz, giving a theoretical resolution of 60~cm. With the help of the MUSIC algorithm for signal separation, the proposed method can detect the vital signs of two people located just 40~cm apart (in terms of their distance from the radar), which exceeds the theoretical range resolution limit.  When subjects were closer than 40~cm, they could be spatially identified, but the vital sign signals interfered, and the results were not accurate. Larger non-vital movements posed a challenge to the radar, and the authors suggested combining MUSIC with micro-Doppler methods for extracting only VS motions.

A more recent method for measuring VS in a multi-target scenario is presented in \cite{XUE2023113715}. The authors utilize a 77~GHz FMCW radar with a 3.5~GHz bandwidth. The authors used Moving Target Indication for stationary clutter suppression, Constant False Alarm Rate Detection for detecting the distance to the target peaks, and K-means clustering for grouping detected peaks into clusters representing people. The developed method was tested in an office scenario with eight persons (four scenarios of two persons each), with distances from one to two meters. It achieved mean absolute errors of 1.22~bpm for BR and 3.62~bpm for HR, and mean accuracies of 94.2\% and 95.1\%, respectively. 

Apart from using specialized radars for VS monitoring, another approach that has recently gained significant attention is Wi-Fi-based sensing. These methods measure the variations in the received Wi-Fi signals to detect the presence of people in the room and estimate BR or HR.

A notable advancement in Wi-Fi sensing is presented in \cite{Guan2023}. The authors examine the use of modified Wi-Fi devices for simultaneously measuring the BR of two individuals. The scenario involves using an antenna array at the receiver, enabling the resolution of two targets. The carrier frequency is 5.32 GHz, and the receiver features four microstrip Yagi antennas. The setup includes an OFDM communication system. An algorithm evaluates the variance of the channel frequency response for each subcarrier. The subcarriers with the highest variance are most sensitive to motion, so only they are used for sensing. The obtained measurements are highly accurate, particularly when the device is located close to the subjects. For distances below 1.5 meters, the system achieves accuracy above 95\% in both single-person and two-person scenarios. Accuracy degrades with distance, more rapidly in the two-target case.

Another significant contribution is presented in \cite{wifi-posture-sleep}, where the authors analyzed human BR and HR during sleep using off-the-shelf Wi-Fi. The setup included an \texttt{802.11n} commercial TP-Link TL-WDR4300 access point transmitting the signal. The system used CSI from 30 subcarriers evenly distributed across the 56 subcarriers of a 20~MHz OFDM channel. For BR estimation, the proposed algorithm achieved absolute errors of less than 0.4~bpm in over 90\% of measurements for AP--device distances ranging from 2 to 7 meters. The system achieved HR estimation errors of less than 4~bpm in more than 90\% of cases. 

In \cite{gu2021realtime}, the authors described \textit{Wital}, a low-cost vital signs monitoring system based on the same commercial Wi-Fi hardware as described in \cite{wifi-posture-sleep}. The system used CSI data for VS measurement in a non-line-of-sight scenario. A scenario was set up by placing a lead sheet between the transmitter and the receiver. The system achieved a mean BR estimation error of 0.498~bpm with 96.89\% accuracy, and a mean HR estimation error of 3.531~bpm with 94.71\% accuracy. 

Despite recent progress in contactless VS monitoring, most studies focus on FMCW radar, especially at 77~GHz, or on Wi-Fi systems working mostly in 2.4 and 5 GHz ISM bands. There is a gap in the examination of the performance of mmWave OFDM-based systems. Such systems are gaining relevance and represent a key spectrum direction for future wireless networks. In this context, our work demonstrates that an OFDM-based mmWave radar can be used for accurate measurement of human VS and can possibly be integrated into future wireless communication systems. 

\section{Measurement Setup}
\label{sec:setup}

\begin{figure}[ht]
\centering
\includegraphics[width=0.95\linewidth]{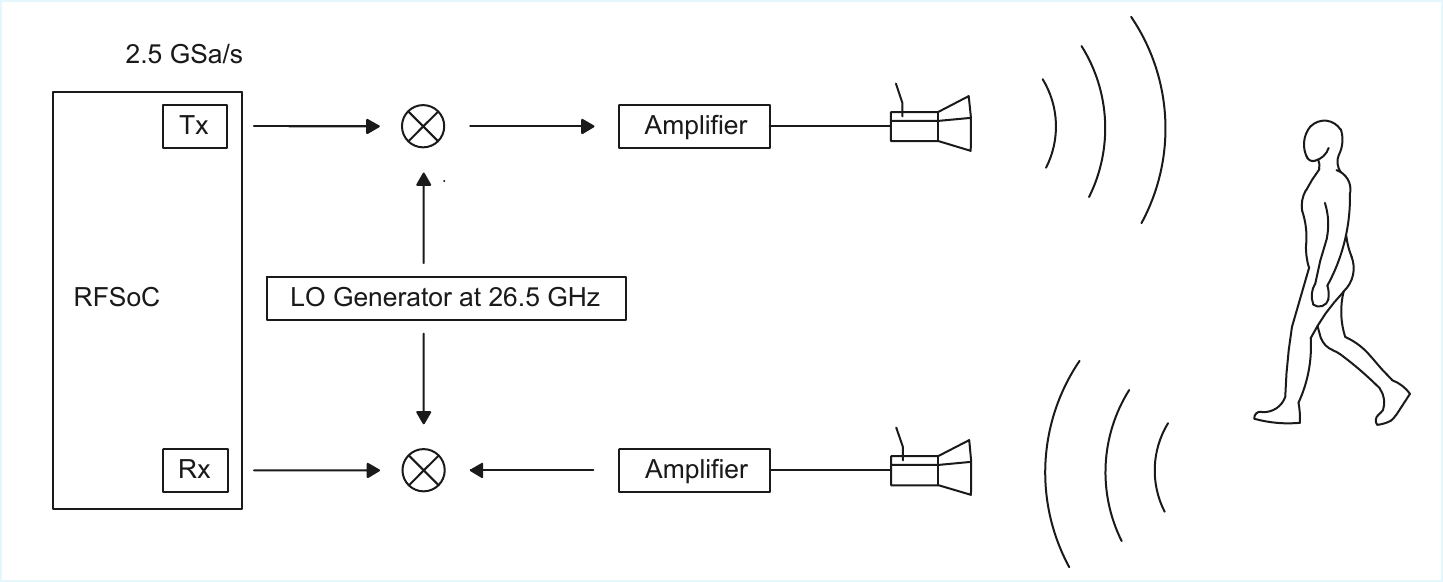}
\caption{The measurement setup.}
\label{fig:setup}
\end{figure}

\begin{figure}[ht]
\centering
\includegraphics[width=0.8\linewidth]{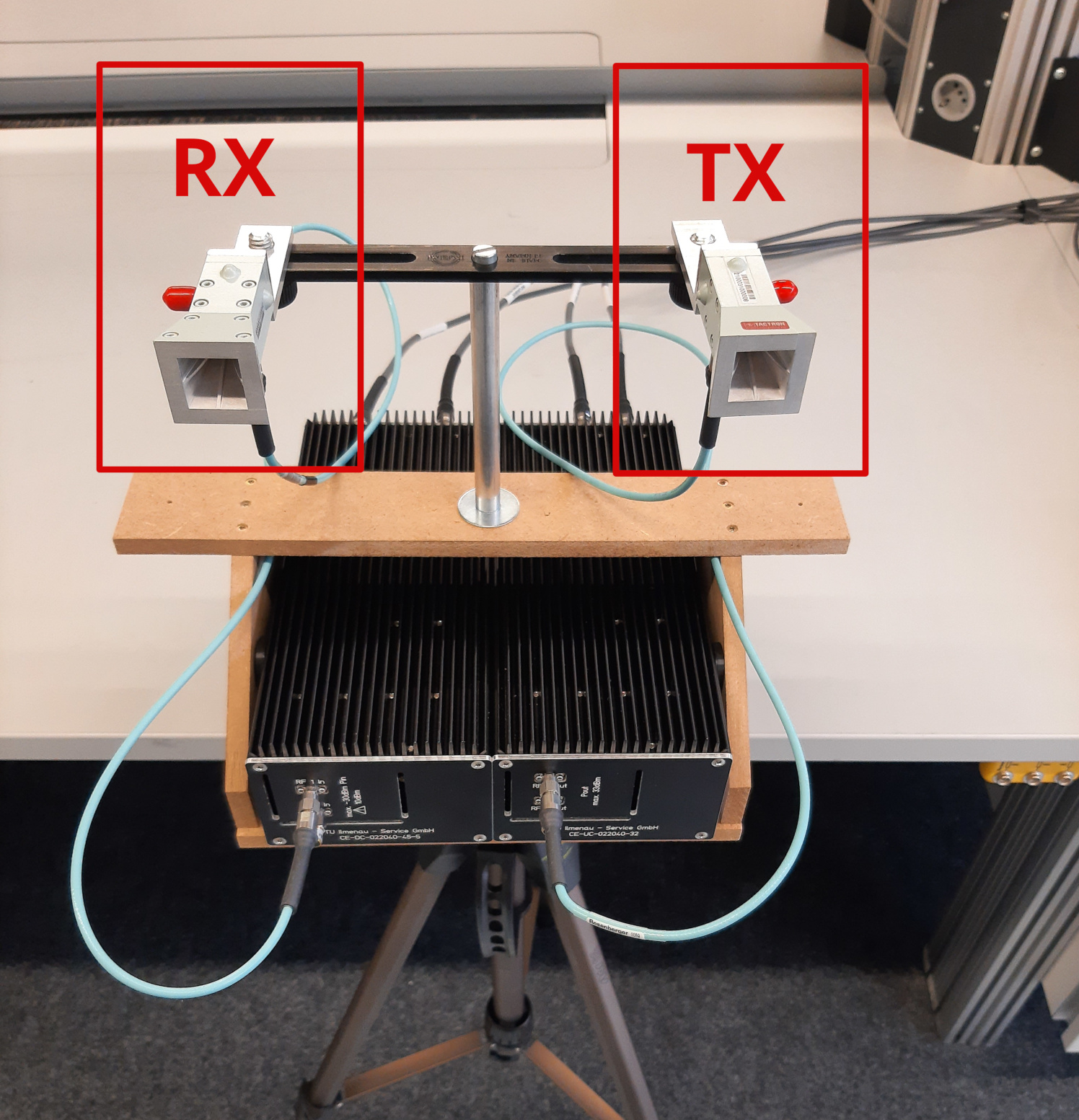}
\caption{A pair of antennas used in the experiments.}
\label{fig:antennas}
\end{figure}

\begin{figure}[ht]
\centering
\includegraphics[width=0.9\linewidth]{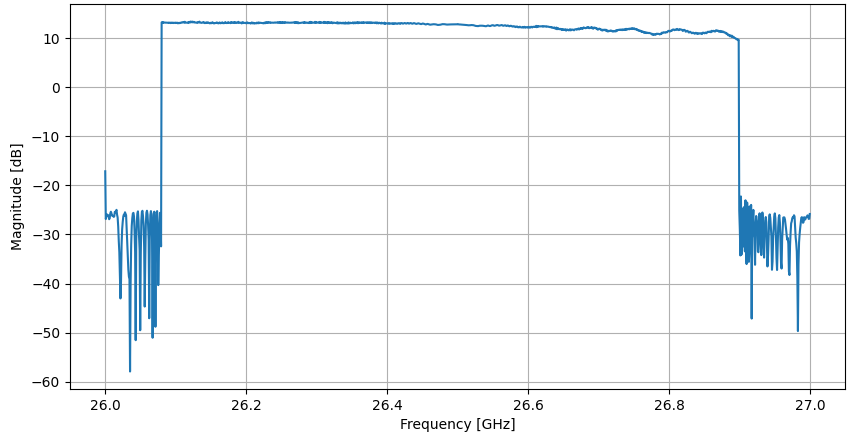}
\caption{Magnitude spectrum of the transmitted OFDM waveform.}
\label{fig:waveform}
\end{figure}

The measurements were taken in the laboratory of the Fraunhofer Institute for Integrated Circuits IIS in Ilmenau, using a software-defined radio (SDR) platform built and maintained by researchers from the hosting institute and the Institute of Information Technology at Technische Universität Ilmenau. The system's architecture is described in \cite{engelhardt2024sdr}. The setup consisted of a bistatic, single-input single-output (SISO) system with two horn antennas, one transmitter (Tx) and one receiver (Rx), specifically, the TACTRON HA-QR-18000-40000-2x2.92F (18–40 GHz, 13 dBi gain at 26 GHz). The diagram of the laboratory setup is presented in Fig. ~\ref{fig:setup}. Fig. ~\ref{fig:antennas} shows the Tx and Rx antennas used in the setup. The waveform was a custom OFDM signal with 1024 subcarriers and a bandwidth of 1 GHz. Its frequency-domain representation is shown in Fig. \ref{fig:waveform}. The subcarriers did not carry any data. Instead, their phases were specifically adjusted to reduce the peak-to-average power ratio, optimizing them for signal processing purposes. The system operated at a carrier frequency of 26.5 GHz, corresponding to a wavelength of approximately 1.13 cm, which is comparable to the amplitudes of body movements caused by HR and respiration. The bandwidth determines the range resolution, which is equal to 15 cm. Therefore, measuring vital signs would not be possible by tracking the changing distance between the antenna and the human body using the time of flight of the signal. The range bin is 7.5 cm, which is too large to capture breathing and heart-induced movements. Each OFDM symbol was sampled 2500 times. The system recorded the reflected signal for multiple OFDM pulses. The recorded data was averaged in slow time because the raw data rate was nearly 20 GB/s. The averaging factor of 100 in the measurements resulted in a lower data rate, allowing for longer continuous patient monitoring. The most essential parameters of the devices used are summarized in Table~\ref{tab:radar_params}.

\begin{table}[ht]
\centering
\caption{Radar System Parameters}
\label{tab:radar_params}
\begin{tabular}{@{}ll@{}}
\toprule
\textbf{Parameter} & \textbf{Value} \\
\midrule
System Type & Bistatic, SISO \\
Center Frequency & 26.5 GHz \\
Wavelength & $\sim$1.13 cm \\
Signal Type & OFDM \\
Number of Subcarriers & 1024 \\
Bandwidth & 1 GHz \\
Sampling Rate ($R$) & 2457.6 MHz \\
OFDM Symbol Length ($P$) & 2500 samples \\
Symbol Duration & 1~$\mu$s \\
Range Resolution & 15 cm \\
\bottomrule
\end{tabular}
\end{table}

\subsection{Measurement Scenarios}
\label{sect:measurement-scenarios}

A set of measurement scenarios was designed to evaluate the radar's performance in different conditions. These scenarios illustrate the potential applications for vital signs monitoring in hospital conditions. Indoor stationary and dynamic scenarios were considered. The experiments aimed to verify the influence of various factors, including the angle and distance between the patient and the radar, the person's position and activity, as well as line-of-sight conditions.  Vital signs were also measured using reference devices to provide ground truth for the radar performance. Holter EKG EDAN SE-2003/201 \cite{edan-holter} and a chest-strap Polar H9 Heart Rate Monitor \cite{polar-h9} were used for pulse monitoring. Breathing rate was measured by extracting the respiratory motion from the patient's video. 

\subsection{A Patient Sitting Still on a Chair}
\label{sect:sitting-still}

\begin{figure}[ht]
\centering
\includegraphics[width=0.95\linewidth]{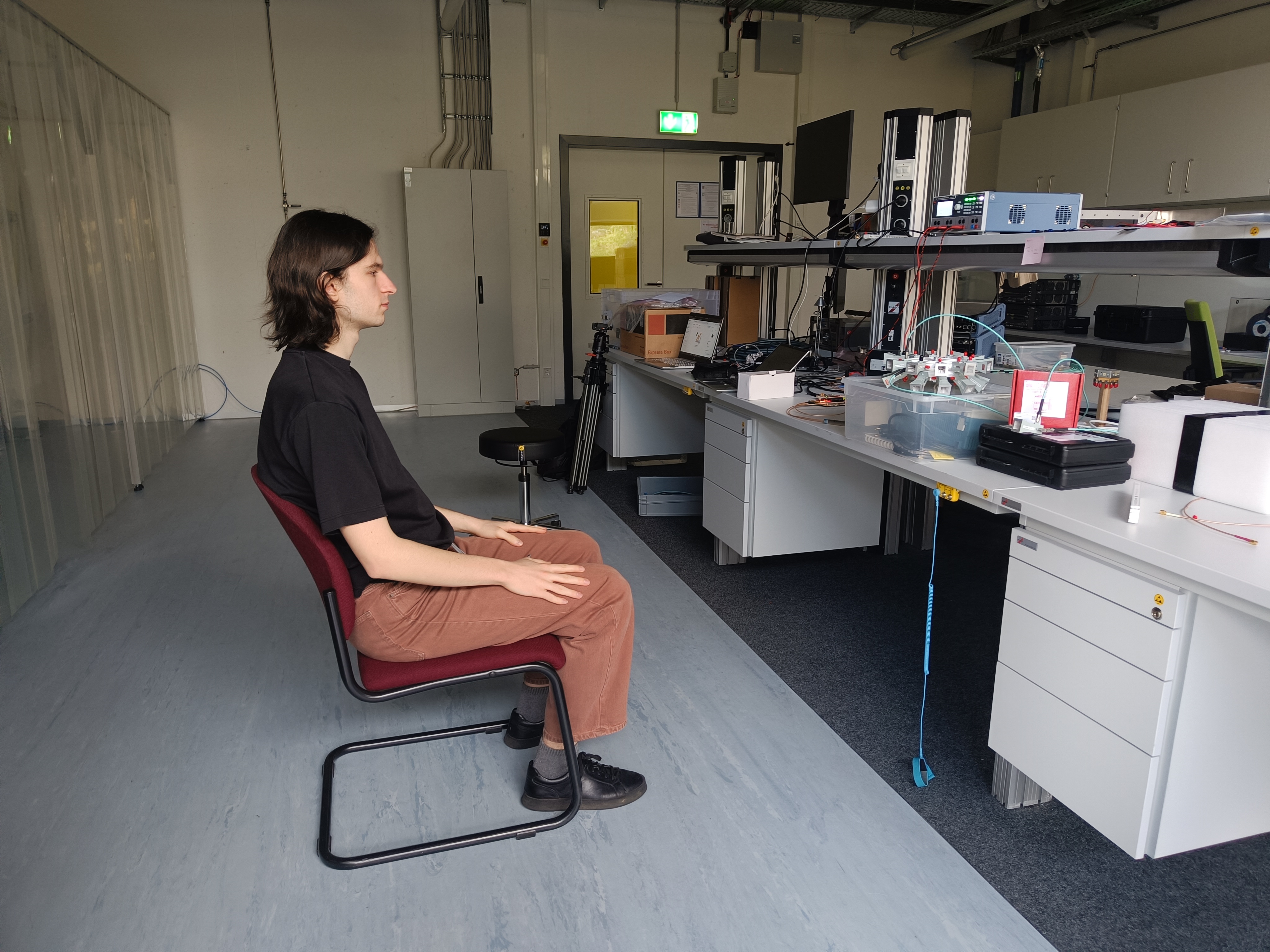}
\caption{The measurement is taken for a sitting person facing the radar.}
\label{fig:patient-sitting}
\end{figure}

\begin{figure}[ht]
\centering
\includegraphics[width=0.95\linewidth]{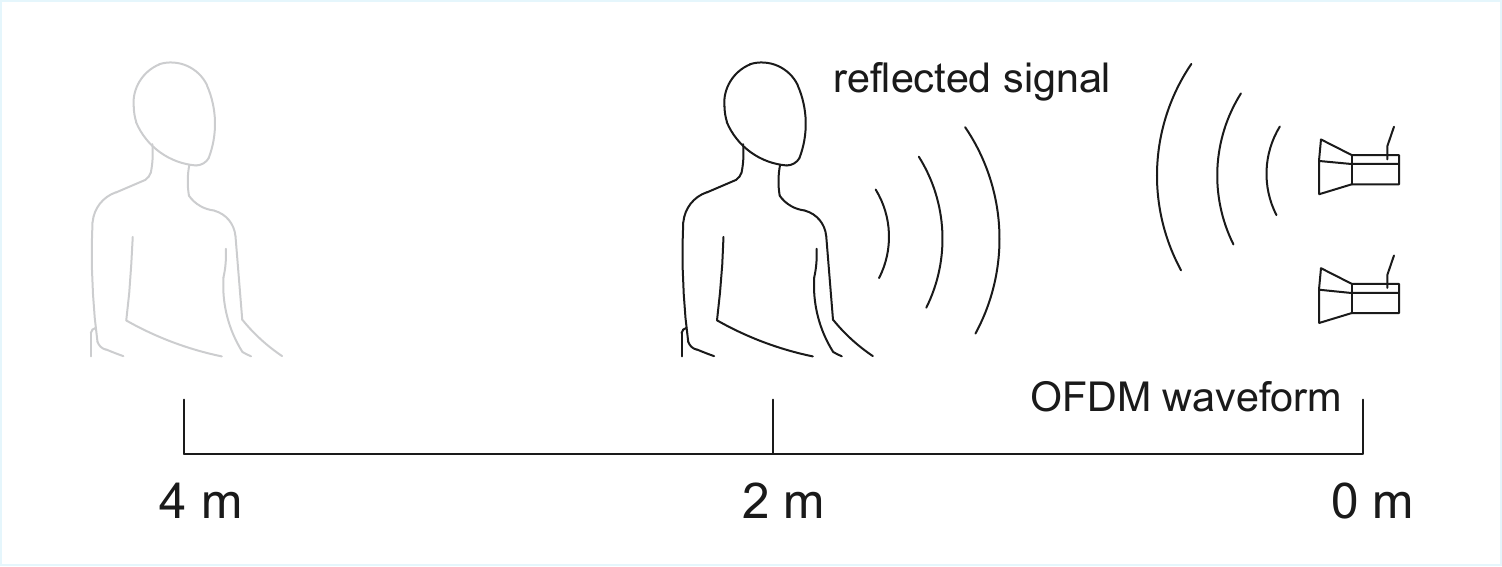}
\caption{Patient VS were measured at four different distances from the antennas.}
\label{fig:sitting-distances}
\end{figure}

The first measurement scenario included a person facing the antennas and sitting still in a chair. The distance was changed from 1 m to 4 meters with a step of one meter. The setup is shown in Figs. \ref{fig:patient-sitting} and \ref{fig:sitting-distances}. 

The examined person was asked not to move nor to look at the radar for the measurement duration, which ranged from 30 seconds to one minute. In some scenarios, the person held their breath to observe only the HR.

\begin{figure}[ht]
\centering
\includegraphics[width=0.8\linewidth]{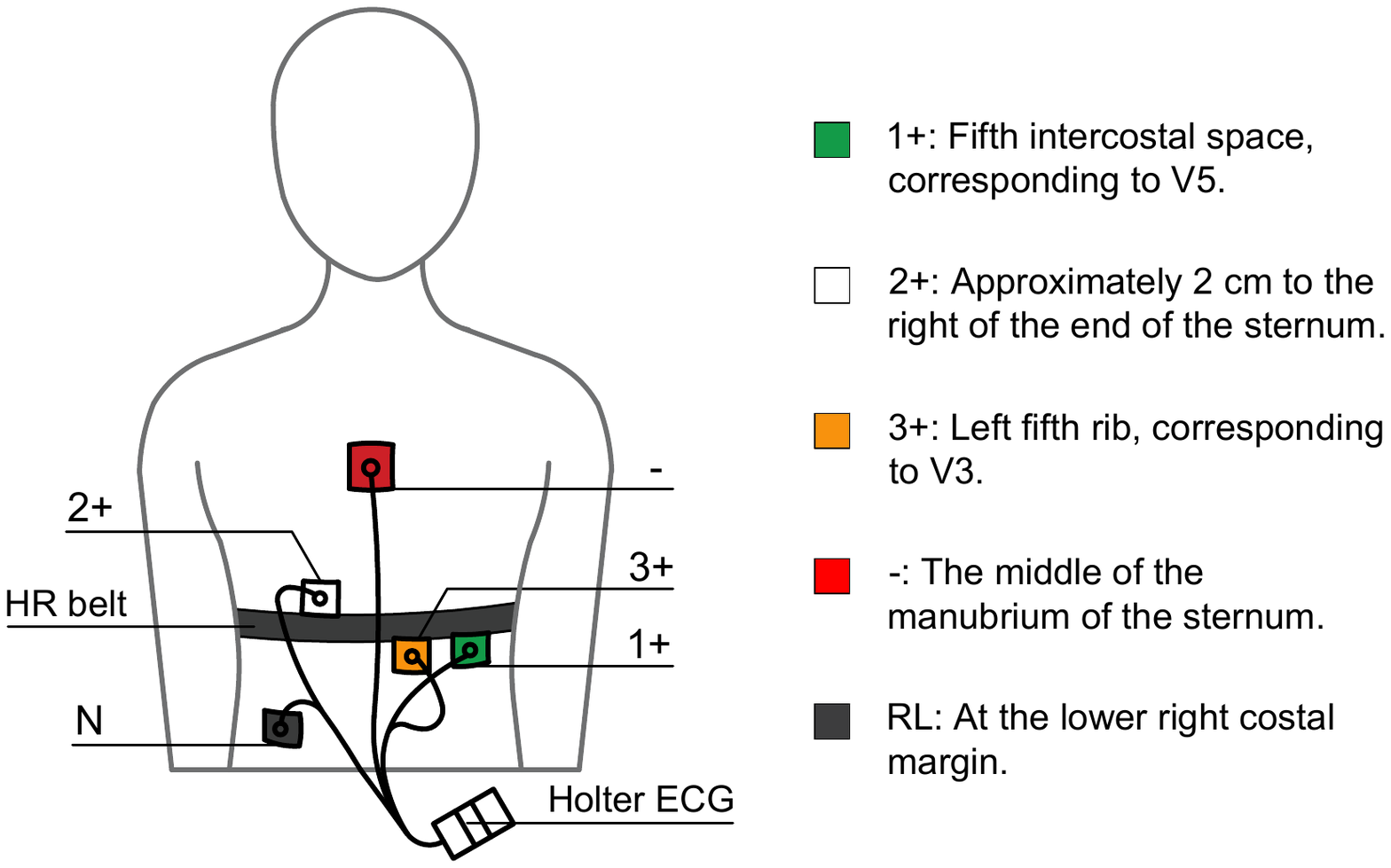}
\caption{A person with a five-electrode Holter ECG device and a HR monitor belt attached.}
\label{fig:electrodes}
\end{figure}

The patient had Holter electrodes attached to measure the HR. A HR belt was also used to give the baseline for our measurements. Measurements were taken with and without a T-shirt, with the latter providing better conditions for VS measurement. The patient was video recorded, as the BR baseline was determined by counting the breaths from the video. A schematic representation of the patient with electrodes and a HR monitor belt attached is presented in Fig. \ref{fig:electrodes}.
To capture the changes in HR, the subject was asked to perform a physical exercise before the measurement. As a result, the HR was expected to decrease until reaching a steady level.

\subsection{Sitting at a Desk}
\label{sect:sitting-at-a-desk}

\begin{figure}[ht]
\centering
\includegraphics[width=0.8\linewidth]{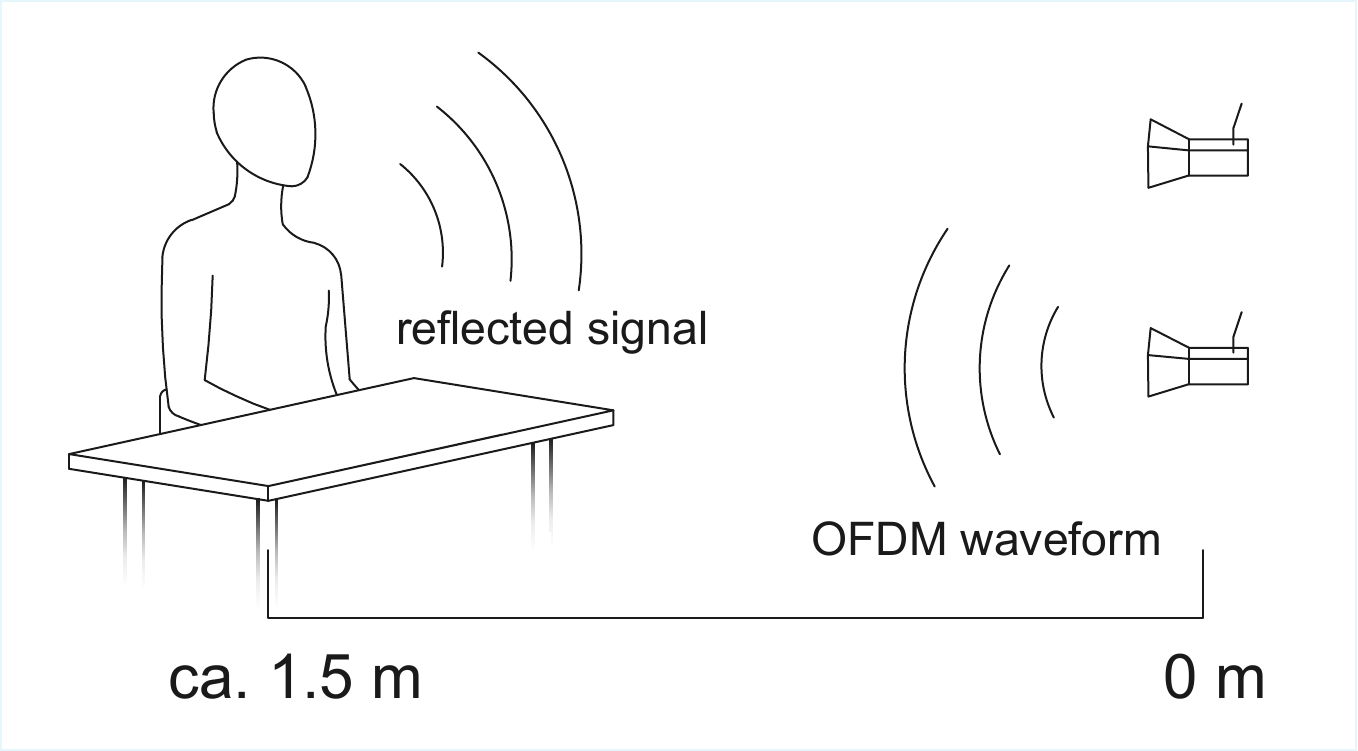}
\caption{The patient sitting at a desk.}
\label{fig:sitting-at-the-desk}
\end{figure}

In the following scenario, the person remained seated at a desk, either remaining still or reading a book while moving their head and arms. The radar was positioned at a slight angle to the patient. The scenario is illustrated in Fig. \ref{fig:sitting-at-the-desk}.

\subsection{Variable Angle Scenario}
\label{sect:variable-angle-scenario}

\begin{figure}[ht]
\centering
\includegraphics[width=0.8\linewidth]{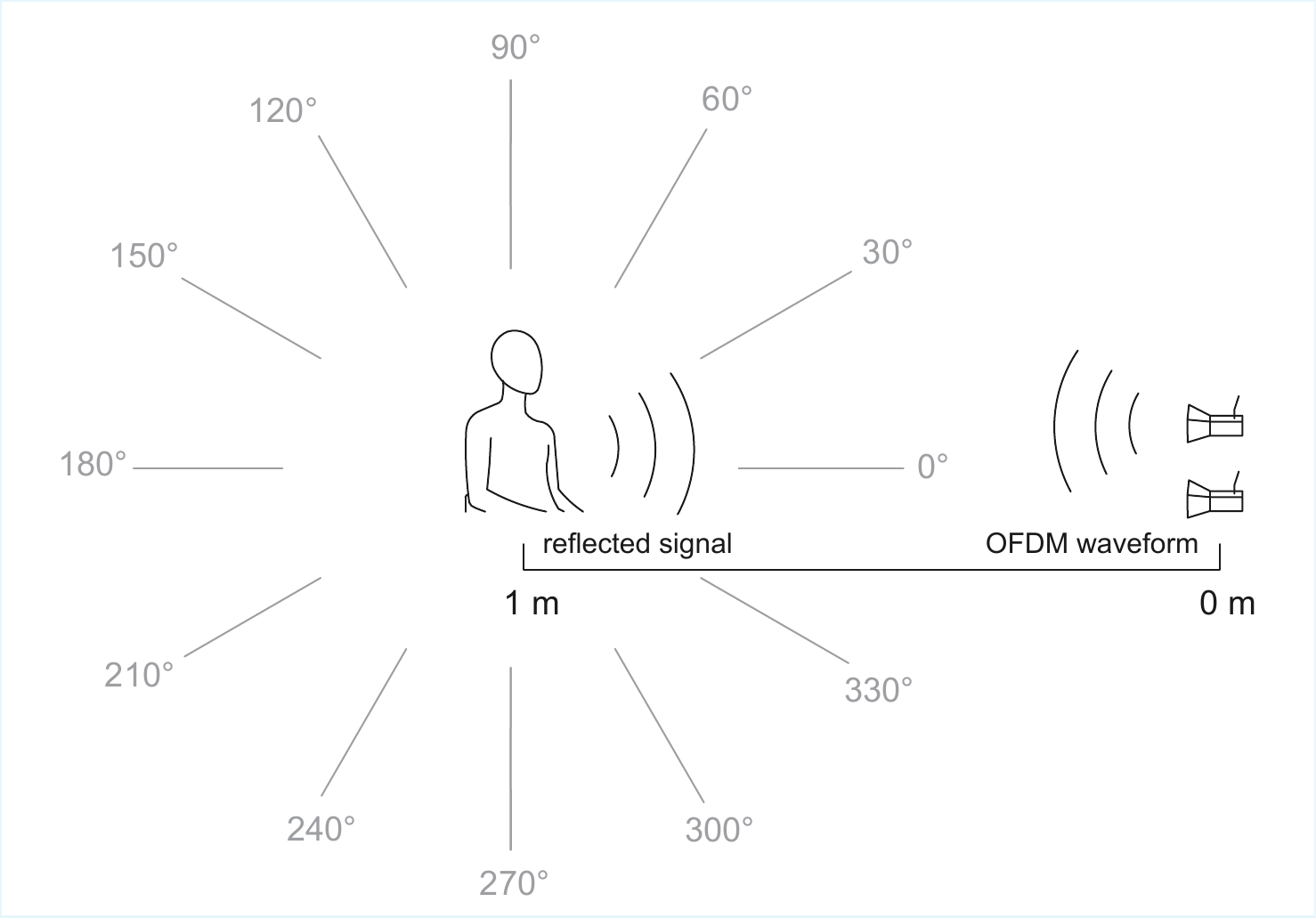}
\caption{A person sitting on a chair at varying angles relative to the radar.}
\label{fig:angles}
\end{figure}

In this scenario, the influence of the orientation between the radar and the subject was investigated. The measured amplitude of the body movement due to respiration and HR is the largest when the movement vector direction is facing the radar. Therefore, it is more challenging to capture vital signs when the patient is oriented in the opposite direction, as the back movements are smaller than those of the chest and front of the neck. The patient was rotated 30 degrees each time to make 12 measurements under different angles, as shown in Fig. \ref{fig:angles}. The distance between the radar and the patient was set to 2 meters. 

The measurements were repeated for five persons, with angles from -90 to +90 degrees, with a step of 30 degrees, and one measurement at -180 degrees. A belt heart monitor was used as a reference. 

Apart from sitting, patients were asked to stand in front of the radar device and remain completely still. A second standing scenario was conducted, in which the patient was instructed to move naturally while standing: making small gestures and moving their head and torso. This scenario simulated more realistic conditions where a person may not remain perfectly still. 

\subsection{Lying Down Position}
\label{sect:lying-down-position}

\begin{figure}[ht]
    \centering
    \includegraphics[width=0.6\linewidth]{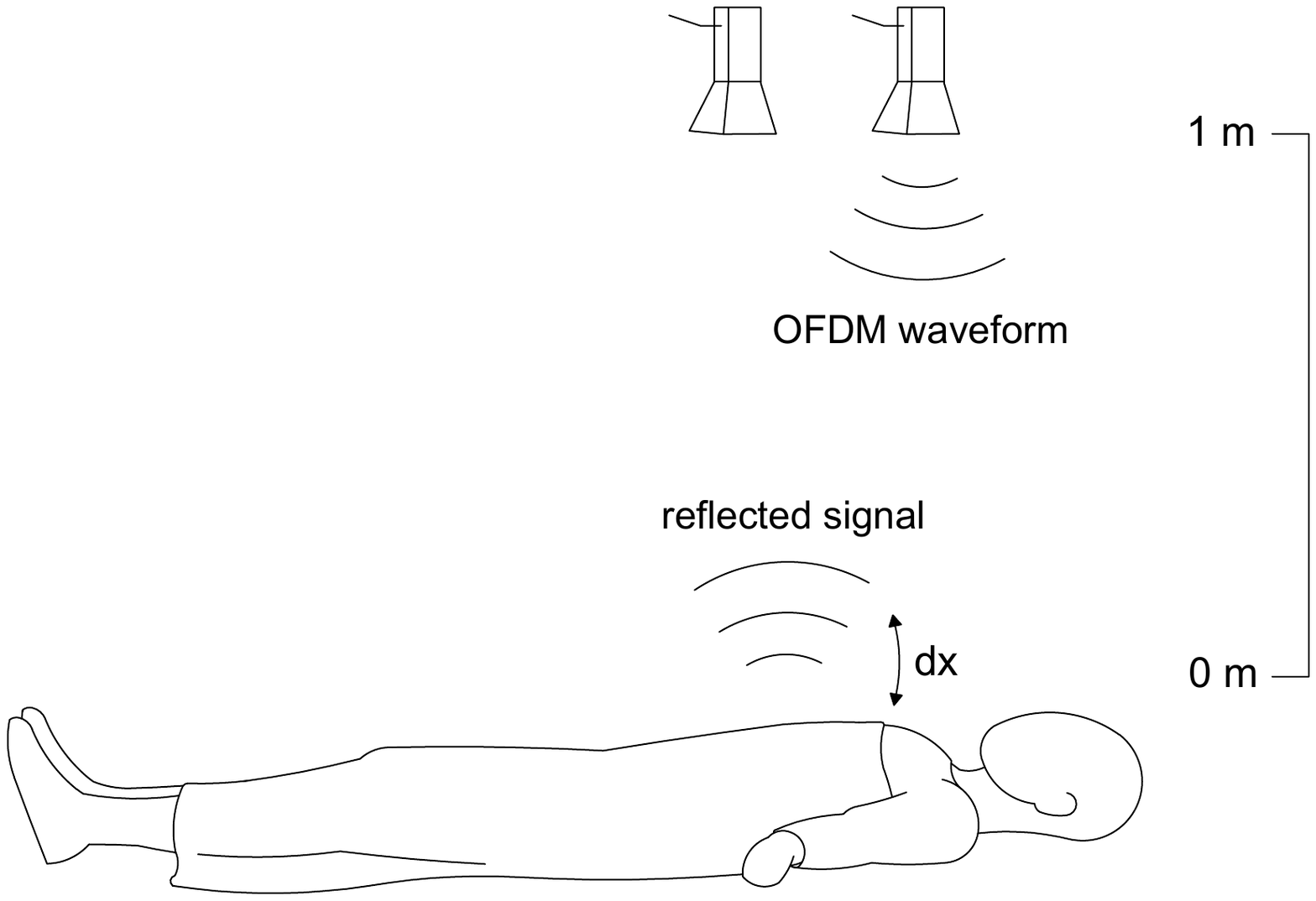}
    \caption{Measurement with the patient lying down.}
    \label{fig:image7}
\end{figure}

In this scenario, illustrated in Fig.~\ref{fig:image7}, the patient was positioned lying down, with the radar mounted above. The person was first measured wearing only a T-shirt. Subsequently, the measurement was repeated with the subject covered by a blanket and wearing a sweatshirt. The scenario aimed to verify the impact of layers of clothing on the radar's ability to measure vital signs. This setup reflects a typical setting in a hospital or elderly care facility, where patients lie in bed under blankets but do not perform rapid movements.

\subsection{Walking Scenario}
\label{sect:walking-scenario}

\begin{figure}[ht]
    \centering
    \includegraphics[width=0.8\linewidth]{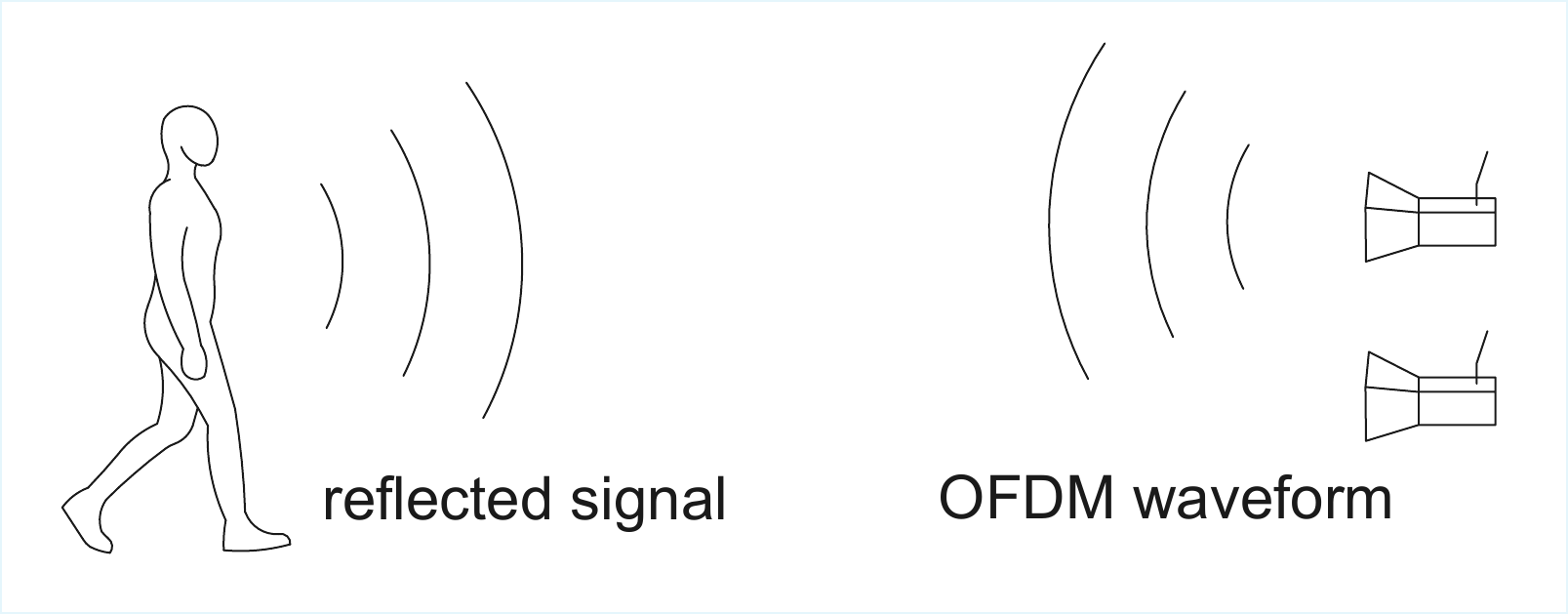}
    \caption{Walking in front of the radar.}
    \label{fig:walking_scenario}
\end{figure}

In this scenario, illustrated in Fig.~\ref{fig:walking_scenario}, the person was asked to walk back and forth in front of the radar. Two measurements were taken: one with a slow walk and a second with a faster pace. This experiment aimed to assess the radar's ability to detect vital signs. At the same time, the person was in motion, which can be challenging due to the overlapping movements of walking, breathing, and the heartbeat. If VS extraction became difficult or impossible, the radar was expected to identify the presence of movement even if VS could not be reliably extracted.

\subsection{Non-Line-of-Sight Conditions}

\begin{figure}[ht]
    \centering
    \includegraphics[width=0.8\linewidth]{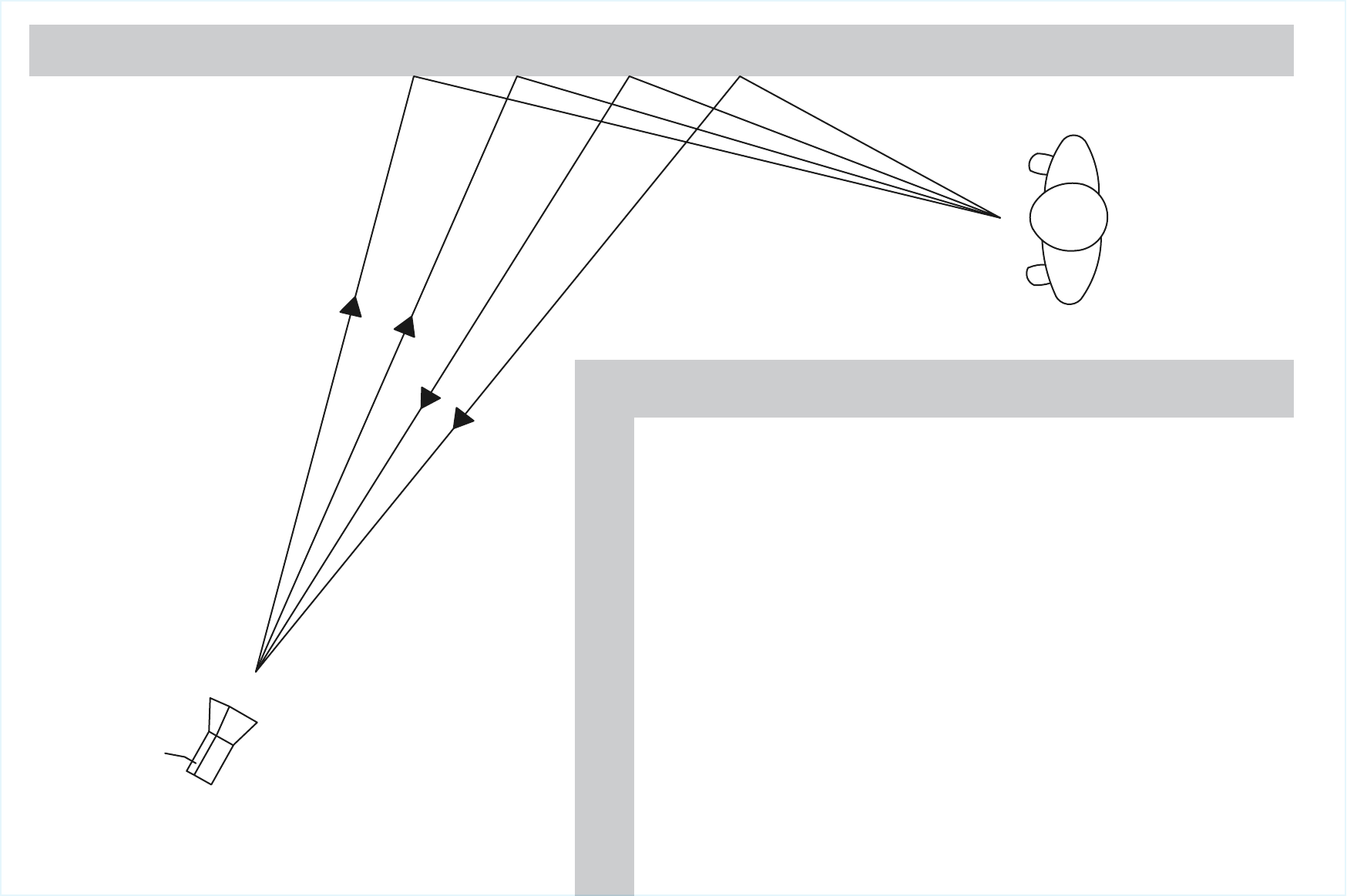}
    \caption{NLOS measurement setup. Person standing behind an obstacle.}
    \label{fig:nlos_scenario}
\end{figure}

In this scenario, the subject was placed behind an obstacle that blocked the direct line of sight, as illustrated in Fig.~\ref{fig:nlos_scenario}. The person stood still and breathed calmly. The aim was to evaluate whether the radar could effectively measure vital signs using only non-line-of-sight (NLOS) components of the reflected signal. This test setup simulates realistic conditions in clinical or home environments, where furniture or walls might obstruct direct visibility.

\subsection{Vital Sign Monitoring of Two Persons}

\begin{figure}[ht]
    \centering
    \includegraphics[width=0.8\linewidth]{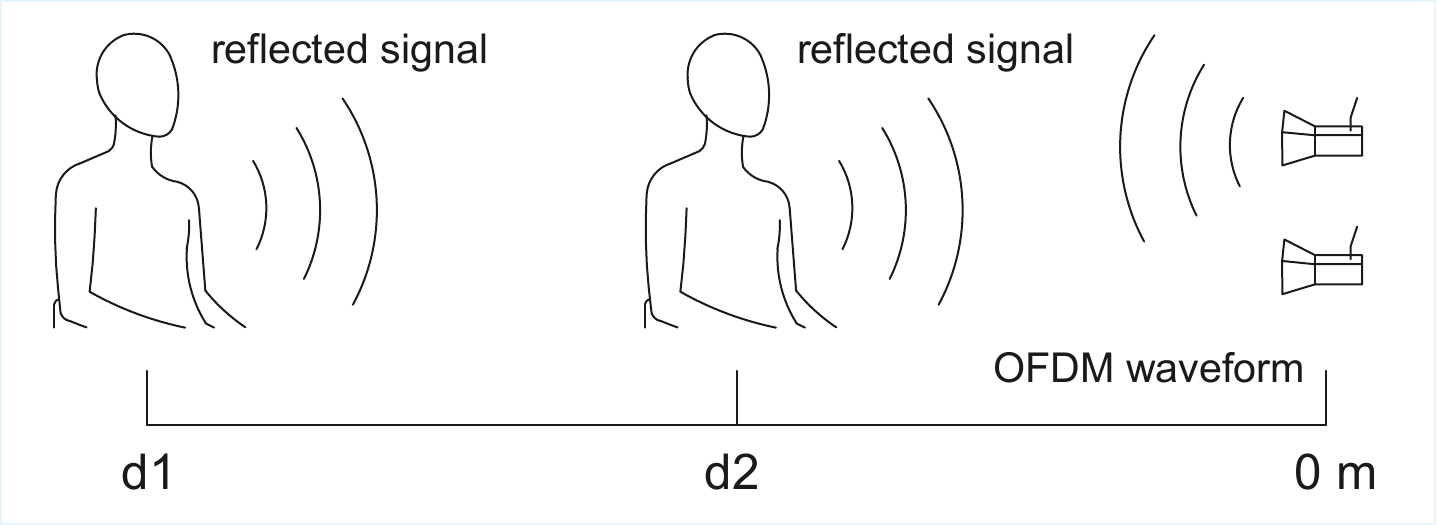}
    \caption{VS monitoring of two persons.}
    \label{fig:two_persons}
\end{figure}

To test the radar's performance in monitoring more than one individual simultaneously, a scenario was set up where two persons were seated at different distances, facing the radar. The person closer to the radar did not cover the radar's view of the further person. The scenario is illustrated in Fig.~\ref{fig:two_persons}.

A summary of all the experiments is provided in Table~\ref{tab:experiment_summary}.

\begin{table*}[ht]
\begin{center}
\caption{Summary of Experimental Scenarios}
\label{tab:experiment_summary}
\begin{tabular}{@{}p{4cm}p{6cm}p{6cm}@{}}
\toprule
\textbf{Scenario} & \textbf{Setup Description} & \textbf{Variables / Conditions} \\
\midrule
Sitting still on a chair & The person sits on a chair, facing the radar at distances from 1~m to 4~m, with a 1~m step. & Distance (1--4~m), Holding breath vs normal breathing \\
\midrule
Sitting at a desk & A person sits at a desk, completely still, or reads a book while moving their head and arms. & Motion (still vs moving arms/head), Radar positioned at a 30° angle \\
\midrule
Variable angle scenario & Patient rotated from --90° to +90° in 30° steps, plus one at --180°. Radar at 2~m. Repeated for five persons. & Rotation angle \\
\midrule
Standing (Still / Natural Movement) & The subject stands in front of the radar, either completely still or performing small natural gestures. & Standing posture, Standing with natural motion \\
\midrule
Lying down position & The patient lies down, radar mounted above. Measured with a T-shirt only, then with a blanket and a sweatshirt. & Clothing layers (T-shirt, blanket, sweatshirt) \\
\midrule
Walking at different speeds & The subject walks back and forth in front of the radar. Two measurements: slow and fast pace. & Walking speed (slow, fast) \\
\midrule
NLOS & The subject stands still behind an obstacle, blocking the direct line of sight. & NLOS conditions \\
\midrule
Two persons & Measuring the vital signs of two patients sitting at different distances from the radar. & \\
\bottomrule
\end{tabular}
\end{center}
\end{table*}

\section{Signal Processing}

Human vital signs, such as breathing and heartbeat, result in periodic movements of the human body, primarily in the thoracic region, neck, and face. During respiration, the expansion and contraction of the lungs cause the chest wall to move. Heart activity generates periodic changes in blood pressure as it ejects blood into the arterial system, creating a pressure wave propagating through the arterial walls. This pressure change causes the pulsation of arteries as they expand and contract rhythmically. These periodic expansions result in a small mechanical displacement of the surrounding tissue and skin. In areas where the arteries are located just under the skin, their movement can induce vibrations on the surface with an amplitude range of 0.2--0.5~mm and a frequency range of 1--1.34~Hz (approximately 60--80 beats per minute)~\cite{Shafiq2014}. The chest surface typically moves within a range of 4--12~mm with a frequency of 0.2--0.34~Hz (approximately 12--20 breaths per minute)~\cite{Shafiq2014}.

In the experiments, a custom-built radar operating at a center frequency of 26.5~GHz was used for data collection. This places the system in the lower part of the mmWave band, where human skin shows significant reflectivity due to its high water content and dielectric properties. When mmWave electromagnetic radiation contacts the person's skin, some of the radiation is reflected, and the rest is absorbed near the skin surface~\cite{Owda2017}. In the lower regions of the mmWave band, for normal incidence—where the wavefront is perpendicular to the skin surface—approximately 40--50\% of the incident electromagnetic wave is reflected at the air–skin interface~\cite{Wu2015}. The actual value depends on the angle of incidence, wave polarization, exact frequency, and water content of the skin.

\subsection{Range Detection}

The signal reflected in the direction of the receiving antenna is measured to extract the target's VS. By continuously monitoring the phase difference between successive radar pulses, it is possible to measure BR and HR. Body movements change the distance that the wave travels between the transmitter, the target, and the receiver. This distance cannot be directly measured based on the time of flight of the signal because the resolution of such a measurement is at least one order of magnitude higher than the measured movements. Therefore, another method was employed for this purpose. The phase of the received signal is calculated for each pulse. Movements of the body cause periodic changes in the received phase, as the distance the wave has to travel varies. By extracting frequency components of the phase variation, one can detect different movements of the human body. Since this study focuses on heart and breathing activity, the analysis explicitly targets low-frequency components associated with respiration and heartbeat. 

\subsection{Digital Signal Processing Pipeline for Vital Sign Extraction}

The raw data captured by the receiver consisted of a stream of samples obtained after the radar signal was down-converted and digitized. Each recorded OFDM pulse lasted 1~$\mu$s and was sampled 2500 times. The radar transmitted these pulses continuously with a pulse repetition interval of 1~$\mu$s, meaning the pulses were sent without gaps. At the beginning of signal processing, the raw data was reshaped into a two-dimensional matrix, where each row corresponded to a single OFDM pulse. As a result, the matrix dimensions are $N \times 2500$, where $N$ represents the total number of pulses received over the measurement period. The first dimension, \textit{slow time}, corresponds to the sequence of pulses transmitted and received over time, capturing the target's activity across successive pulses. The second dimension represents \textit{fast-time} samples within a pulse, which are used to measure the range of the target.

\begin{figure}[t]
    \centering
    \includegraphics[width=0.5\textheight, angle=270]{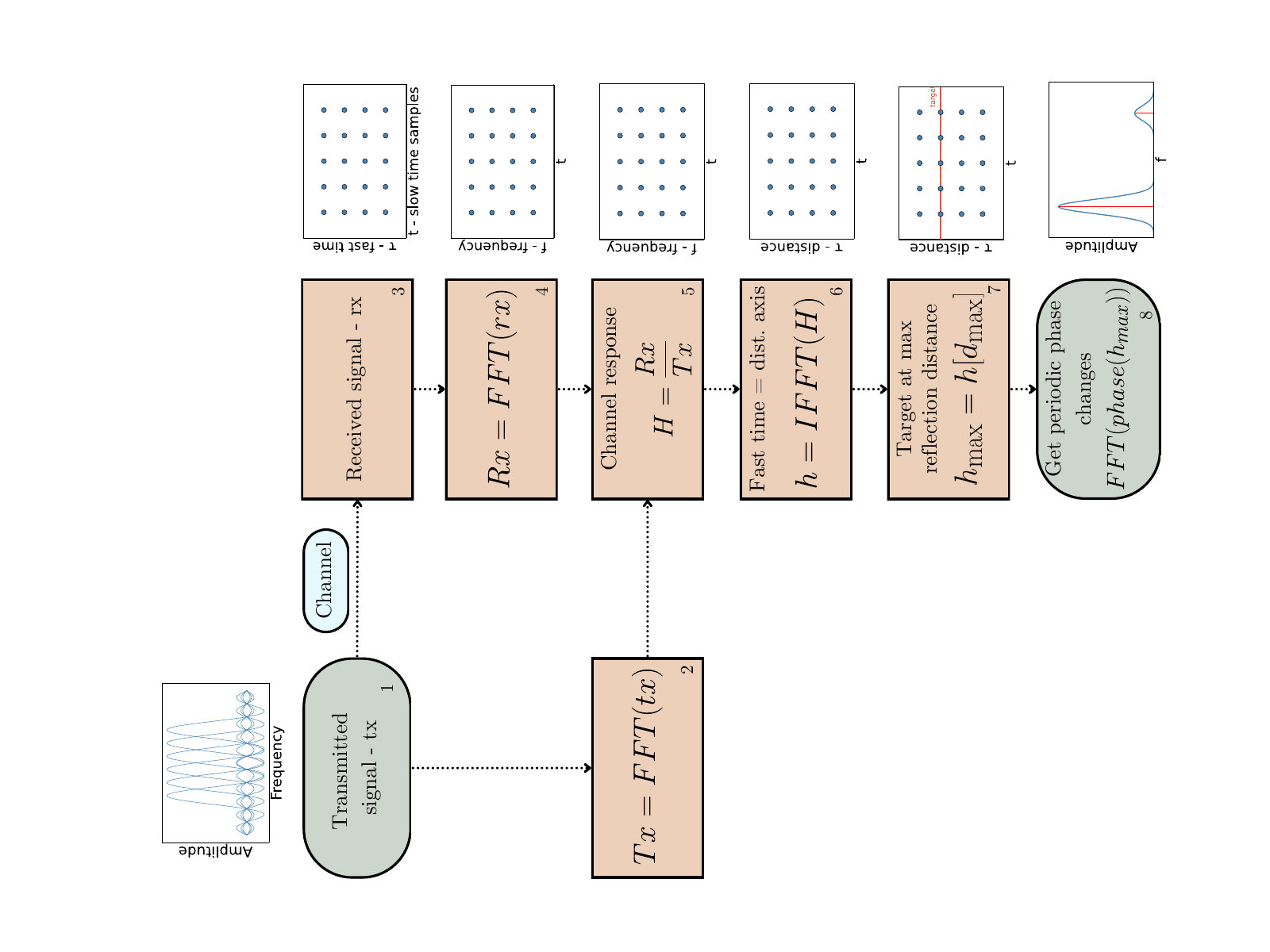}
    \caption{Signal processing pipeline used for extracting vital signs from radar data.}
    \label{fig:pipeline}
\end{figure}

The signal processing pipeline is shown in Fig.~\ref{fig:pipeline}. In the first step, the OFDM-modulated signal is transmitted in the direction of a person. The transmitted signal is transformed into the frequency domain via the Fast Fourier Transform (FFT) in the second step. The transmitted signal propagates through the channel, and some of it is reflected from the human body and received by the second antenna. In the third step, the received signal is transformed by the FFT. To calculate the channel transfer function $H$, the spectrum of the received signal is multiplied by the inverse of the transmitted signal spectrum. The matrix $H$ contains the channel's response across successive pulses (slow time). In the sixth step, an inverse FFT (IFFT) calculated over the frequency axis is used to convert the channel transfer function to the channel impulse response. In the seventh step, the signal matrix is processed to extract only the signal reflected from the subject. To achieve this, the distance between the person and the antenna is calculated based on the propagation delay corresponding to the strongest return signal, averaged over consecutive pulses. The fast-time axis is scaled by multiplying by the radar's range resolution and adjusted by subtracting the propagation distance of the coaxial cable connecting the antennas to the processing unit.

\begin{figure}[ht]
    \centering
    \includegraphics[width=0.45\textwidth]{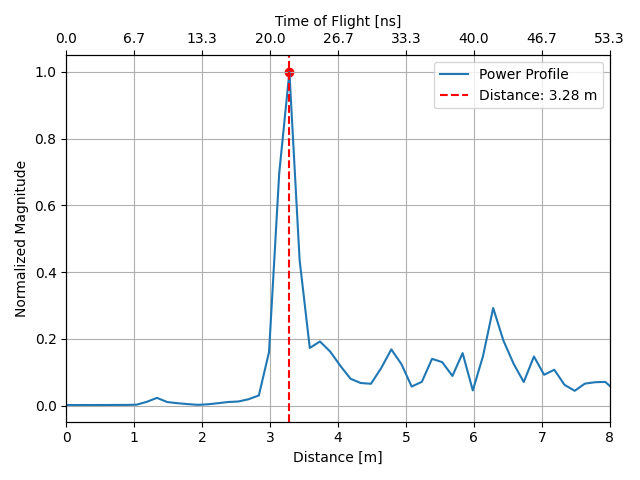}
    \caption{Measured distance to a single person.}
    \label{fig:distance_one}
\end{figure}

\begin{figure}[ht]
    \centering
    \includegraphics[width=0.45\textwidth]{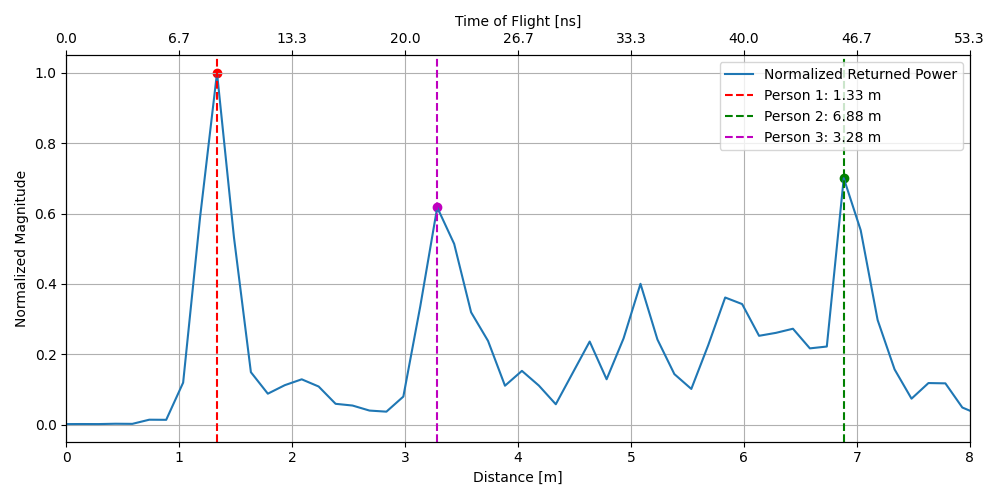}
    \caption{Measured distances to three persons.}
    \label{fig:distance_two}
\end{figure}

The resulting signals with measured distances to one and two persons are shown in Figures~\ref{fig:distance_one} and~\ref{fig:distance_two}, respectively.

\begin{figure}[ht]
    \centering
    \includegraphics[width=\linewidth]{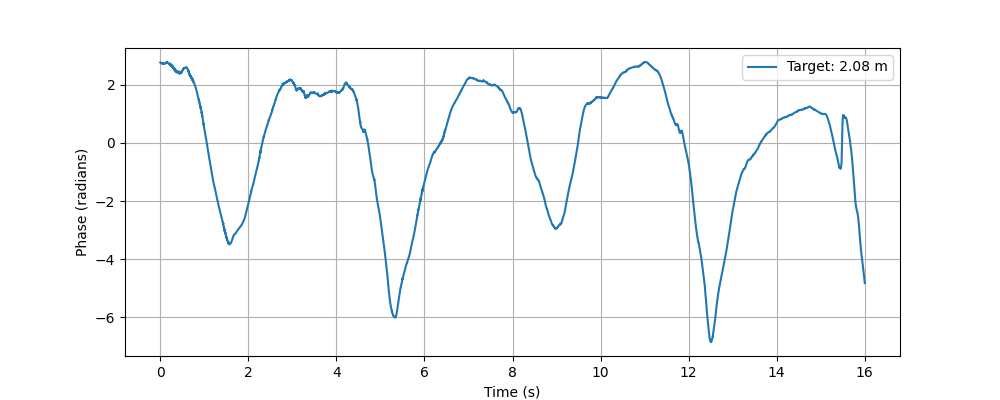}
    \caption{Raw phase variation of the reflected radar signal over time at the measured distance to the person, four full breathing cycles.}
    \label{fig:phase_variation}
\end{figure}

\subsection{Phase Analysis}

After isolating the signal reflected from the subject, the next step is to analyze the phase variation at that distance. As the phase of a complex signal sample is between $-\pi$ and $\pi$, the raw phase of the received signal often contains discontinuities when it crosses these boundaries. Phase unwrapping is applied to remove the cuts in the phase signal. 

The raw phase variation in time is shown in Fig.~\ref{fig:phase_variation}. It corresponds to the changing distance between the person's body and the antennas. The figure shows four complete breathing cycles, each lasting approximately four seconds. The amplitude of each cycle is around six radians, which means that the wave traveled three periods more when the distance was maximal compared to when it was minimal. The wavelength at a frequency of 26.5~GHz equals 1.13~cm. Therefore, three periods correspond to approximately 3.4~cm of breathing movement amplitude. The phase variation caused by the human heart has a much lower amplitude and, therefore, is not easily visible on the raw phase variation plot.

\begin{figure}[ht]
    \centering
    \includegraphics[width=\linewidth]{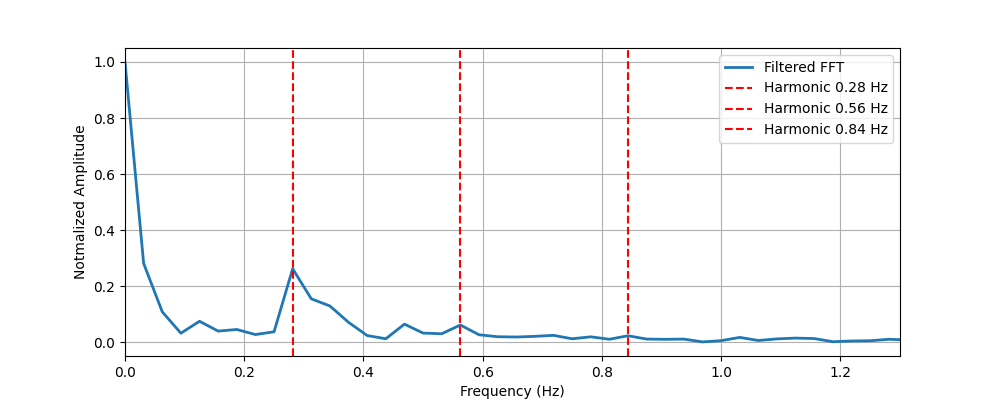}
    \caption{Frequency components of the phase variations with breathing and its harmonics marked.}
    \label{fig:fft_spectrum}
\end{figure}

In the next step, the phase signal is filtered to retain only the frequency components corresponding to BR and HR. Since the breathing frequency typically falls within the range of 0.2–0.34~Hz and HR lies between 1 and 1.34~Hz, a combination of low-pass and high-pass filters is applied to eliminate components outside these ranges, with some margin considered.

After filtering, the signal is transformed into the frequency domain using the FFT. The resulting spectrum is shown in Fig.~\ref{fig:fft_spectrum}. The primary BR frequency is observed at 0.26~Hz. However, because the breathing signal is not a perfect sinusoid, it produces harmonics that appear as multiples of the primary frequency. These harmonics can interfere with the HR signal if they overlap in frequency.

\begin{figure}[ht]
    \centering
    \includegraphics[width=\linewidth]{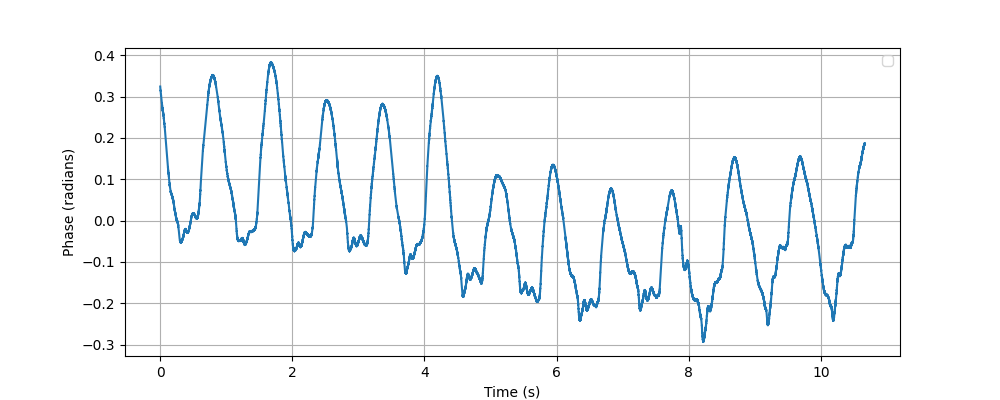}
    \caption{Raw phase variation during a breath-hold.}
    \label{fig:hr_phase}
\end{figure}

\begin{figure}[ht]
    \centering
    \includegraphics[width=\linewidth]{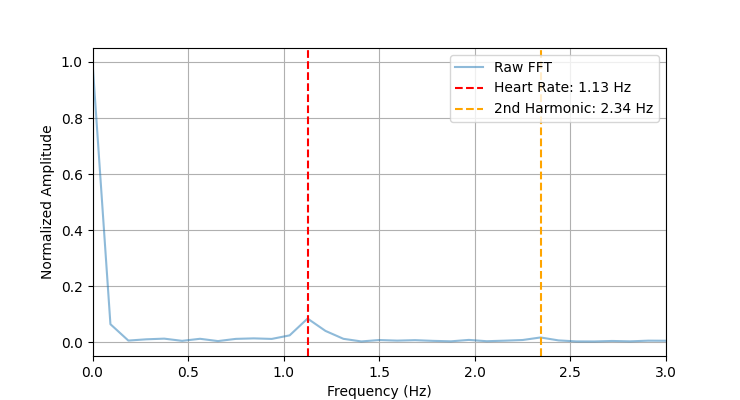}
    \caption{Frequency-domain representation of the signal during breath-hold.}
    \label{fig:hr_fft}
\end{figure}

\subsection{Isolating Heart Rate Signal}

The HR frequency is not always clearly visible when displayed together with the breathing signal due to the amplitude difference. Therefore, the signals were often processed separately to focus on the desired HR frequency range only. 

To demonstrate how the phase changes due to heart activity, the subject was asked to stop breathing during the measurement to isolate the breathing signal. The raw phase variation is shown in Fig.~\ref{fig:hr_phase}. The pulsation amplitude is approximately 0.4 radians, corresponding to a movement amplitude of around 2.26~mm. 

The frequency representation of the HR signal is shown in Figure~\ref{fig:hr_fft}.

\section{Results}

\begin{table*}[t]
    \centering
    \caption{Radar measurements of HR and BR compared with the reference.}
    \label{tab:vital_signs_comparison}
    \begin{tabular}{lcccc}
        \toprule
        \textbf{Scenario} & \textbf{Radar BR [bpm]} & \textbf{Reference BR [bpm]} & \textbf{Radar HR [bpm]} & \textbf{Reference HR [bpm]} \\
        \midrule
        Intermittent breathing (sitting at a desk)      & 20 & 23 & 73.1 & 73.2 \\
        Holding breath                                   & -- & -- & 75.8 & 74.4 \\
        Sitting still                                    & 18 & 19 & 73.8 & 77.4 \\
        Moving while sitting at a desk                   & 15 & 18 & --   & --   \\
        Lying with T-shirt only                          & 14 & 14 & 60.0 & 59.0 \\
        Lying with sweatshirt and thick blanket          & 14 & 14 & 53.4 & 56.3 \\
        NLOS                                             & 17 & 16 & --   & --   \\
        2m no sweatshirt                                 & 17 & 16 & 67.8 & 68.0 \\
        2m with sweatshirt                               & 17 & 17 & 78.6 & 74.7 \\
        4m no sweatshirt                                 & 20 & 14 & 67.8 & 75.3 \\
        4m with sweatshirt                               & 20 & 16 & 67.8 & 70.9 \\
        \bottomrule
    \end{tabular}
\end{table*}

\begin{table*}[ht]
    \centering
    \caption{Effect of angle on radar-based BR and HR measurements at 2 meters distance.}
    \label{tab:angle_effect}
    \begin{tabular}{rcccc}
        \toprule
        \textbf{Angle (°)} & \textbf{Radar BR [bpm]} & \textbf{Reference BR [bpm]} & \textbf{Radar HR [bpm]} & \textbf{Reference HR [bpm]} \\
        \midrule
         0    & 17 & 18 & 78.6 & 73.2 \\
        -30   & 13 & 16 & 75.3 & 73.8 \\
        -60   & 17 & 16 & --   & --   \\
        -90   & 15 & 18 & --   & --   \\
       -180   & 13 & 16 & --   & --   \\
         90   & 15 & 15 & 81.0 & 81.1 \\
         60   & 15 & 15 & 69.6 & 78.7 \\
         30   & 15 & 18 & 70.8 & 73.8 \\
        \bottomrule
    \end{tabular}
\end{table*}

\subsection{Patient Facing the Radar}

The results are generally accurate, although occasional artifacts can be observed. As shown in Table~\ref{tab:vital_signs_comparison}, increased distance negatively impacts the precision of both BR and HR estimations. Wearing an additional layer of thick clothing, such as a sweatshirt, does not significantly degrade the measurement quality.

\subsection{Changing Angle to the Radar.}

The person's orientation relative to the radar significantly affects the accuracy of vital sign estimation. The most reliable results are achieved at angles up to $\pm60^\circ$. At greater angles, such as $\pm90^\circ$, even BR estimation becomes challenging, as the body moves primarily in a direction perpendicular to the radar. This results in minimal phase variation, reducing the radar’s ability to detect chest movements.

\subsection{Walking}

The primary challenge is that as the person moves, their position in the range domain shifts continuously. The first requirement would be a dynamic range-bin tracking algorithm that combines the signal from different range bins. However, the movements of the body overlap with the VS, making the task challenging. Accurate extraction of vital signs under such conditions remains an open research problem.

\subsection{Non-Line-of-Sight Conditions}

\begin{figure}[ht]
    \centering
    \includegraphics[width=\linewidth]{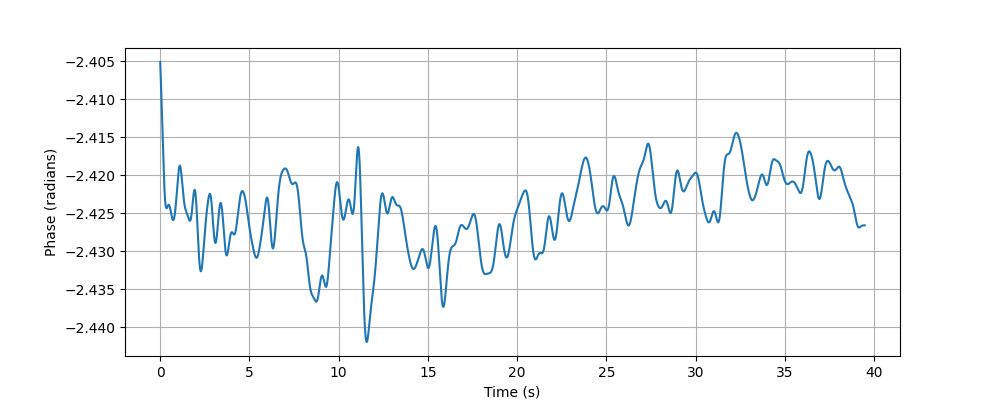}
    \caption{Phase variation for a person behind an obstacle.}
    \label{fig:nlos-phase}
\end{figure}

As shown in Fig.~\ref{fig:nlos-phase}, the amplitude of the phase signal is noticeably reduced compared to line-of-sight scenarios. While the BR can still be estimated, the HR signal is too weak to be reliably extracted.

The results of this measurement are included in Table~\ref{tab:vital_signs_comparison}.

\subsection{Sitting at a Desk}

\begin{figure}[ht]
    \centering
    \includegraphics[width=\linewidth]{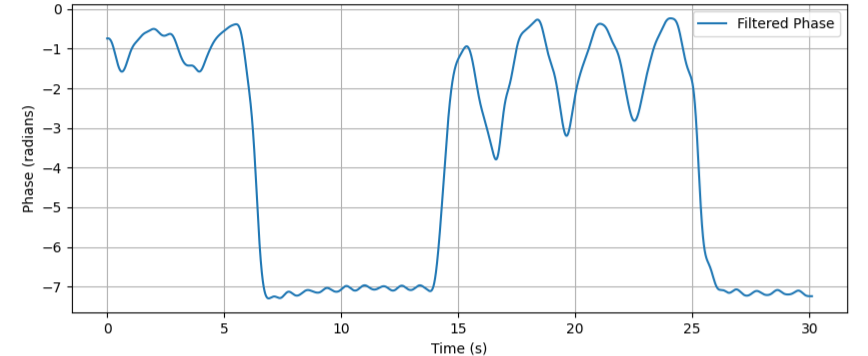}
    \caption{Person periodically holding their breath.}
    \label{fig:intermitted-breathing}
\end{figure}

When a person is seated at a desk and remains stationary, the radar system can reliably detect breathing and HR. The most significant source of inaccuracy in HR measurements is the presence of breathing harmonics. In one of the tested scenarios, the person was instructed to hold their breath periodically to simulate irregular respiration. Fig.~\ref{fig:intermitted-breathing} shows the phase changes of the received signal. Periodical variations with higher amplitude are caused by breathing. Faster changes with lower amplitude are also clearly visible when the person holds their breath.

\begin{figure}[ht]
    \centering
    \includegraphics[width=\linewidth]{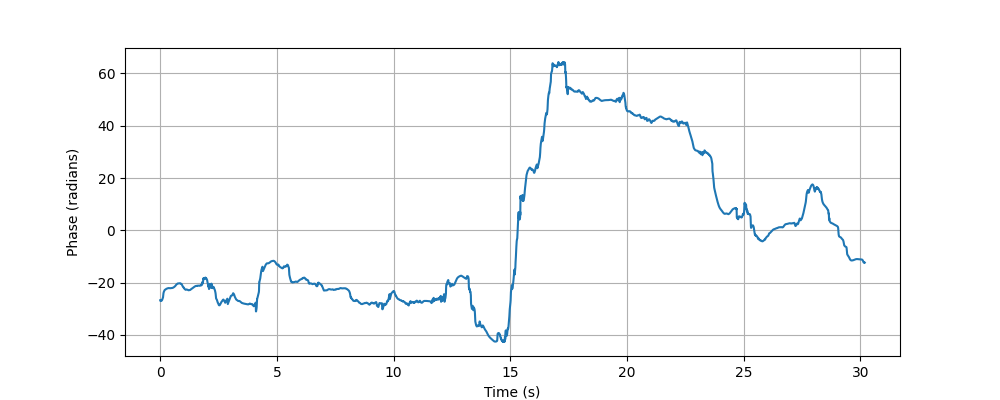}
    \caption{Phase variation of a person moving while seating at a desk. Breathing extracted, HR not distinguishable.}
    \label{fig:moving-at-a-desk}
\end{figure}

When the person was moving at a desk, the phase variation caused by VS became much less visible, as shown in Fig.~\ref{fig:moving-at-a-desk}. 

\subsection{Lying Patient}
Results for the two examined scenarios are presented in Table~\ref{tab:vital_signs_comparison}. BR and BR are measured accurately, with only a slight difference between the radar and reference measurements. This is because the patient remains naturally still while lying in bed, so other body movements do not interfere with the vital signs. Even when the patient is covered with a blanket and wears a sweatshirt, the measurement quality is not significantly affected.

\subsection{Vital Sign Monitoring of Two Persons}

In the first stage of processing, the radar system successfully detected the distances to both participants by identifying peaks in the channel impulse response. Based on these distances, the corresponding range bins with the highest reflected power were selected for further processing.

The radar accurately estimated the BRs of both individuals, as presented in Table~\ref{tab:two_person_measurement}. The values of HR are identical for both patients due to coincidence, as the same frequency bin had the maximum amplitude in both cases. 

\begin{table*}[htbp]
\centering
\caption{Results for two persons monitored simultaneously}
\label{tab:two_person_measurement}
\begin{tabular}{lcccc}
\hline
\textbf{Participant (Distance)} & \textbf{Radar BR [bpm]} & \textbf{Reference BR [bpm]} & \textbf{Radar HR [bpm]} & \textbf{Reference HR [bpm]} \\
\hline
Person 1 (1.6 m)  & 16 & 16 & 71.4 & 73.6 \\
Person 2 (3.44 m) & 18 & 19 & 71.4 & 87.0 \\
\hline
\end{tabular}
\end{table*}

\begin{figure}[H]
    \centering
    \includegraphics[width=\linewidth]{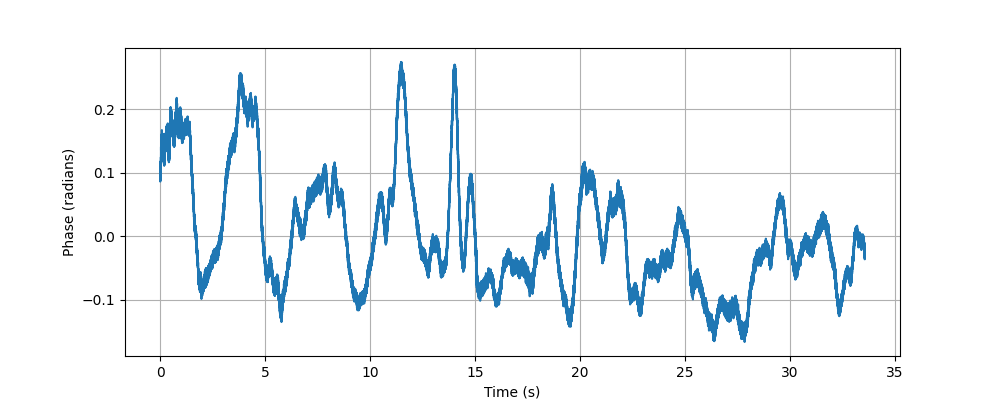}
    \caption{Raw phase variation of Person~2 (d = 3.44~m).}
    \label{fig:p2_phase}
\end{figure}

\begin{figure}[ht]
    \centering
    \includegraphics[width=\linewidth]{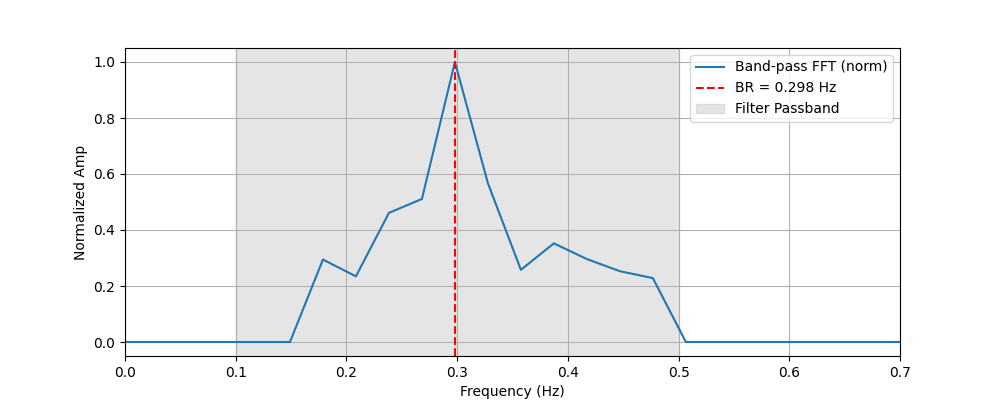}
    \caption{FFT of the phase signal (d = 3.44~m), filtered for breathing frequencies.}
    \label{fig:br_phase}
\end{figure}

\begin{figure}[H]
    \centering
    \includegraphics[width=\linewidth]{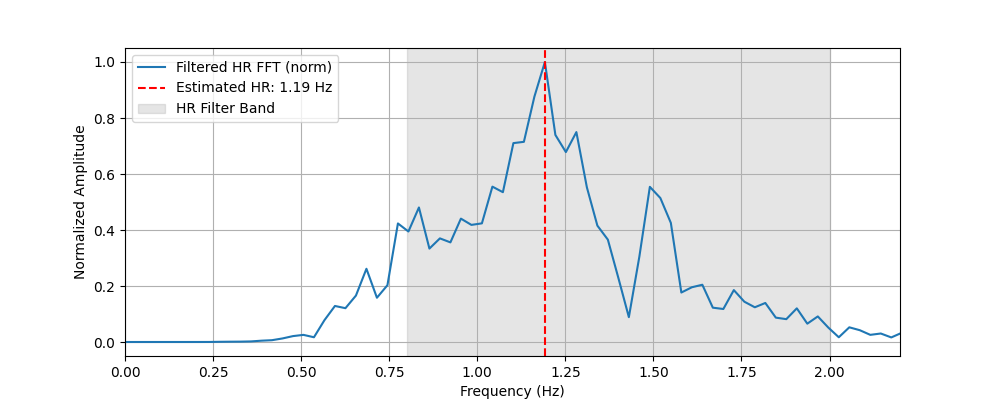}
    \caption{FFT of the phase signal (d = 3.44~m), filtered for HR frequencies. The radar incorrectly identifies a breathing harmonic as HR. The true HR is observed at 1.5~Hz based on Holter reference data.}
    \label{fig:hr2_phase}
\end{figure}

Figures~\ref{fig:p2_phase} through \ref{fig:hr2_phase} illustrate the results obtained for Person~2, positioned at a distance of 3.44 meters from the radar. The measurements for the individual seated closer to the radar were analogous, except that the amplitude of the phase variation was approximately 2 radians, which is around five times larger than that observed for Person~2. Fig.~\ref{fig:p2_phase} shows the unwrapped phase variation over time before filtering. Fig.~\ref{fig:br_phase} presents the FFT of the phase signal filtered for the breathing frequencies. Finally, Fig.~\ref{fig:hr2_phase} displays the FFT filtered for HR frequencies. The algorithm misinterpreted the third harmonic of the breathing signal as the HR. The actual HR, measured by the Holter reference device, was located at approximately 1.5 Hz. In the picture, there is a peak at that frequency, but its amplitude is lower than that of the breathing's harmonic. For the person sitting closer to the radar, the HR was measured correctly.

\subsection{Reducing the Number of Subcarriers}

To evaluate how bandwidth influences the accuracy of the results, the VS were measured using a subset of transmitted subcarriers. In the setup, each subcarrier has a bandwidth of approximately 1~MHz. The numbers of selected subcarriers were 10, 40, and 1024, changing the used bandwidth while keeping the central frequency fixed. 

\begin{figure}[H]
    \centering
    \includegraphics[width=0.5\textwidth]{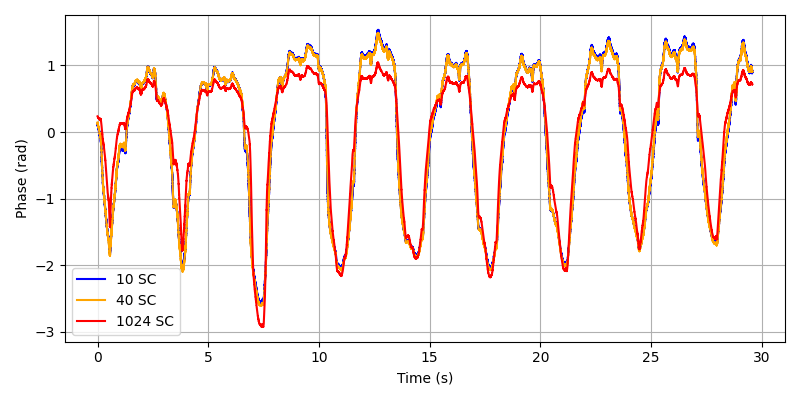}
    \caption{Unwrapped phase for different numbers of subcarriers. Sitting at a desk scenario.}
    \label{fig:compare_phase}
\end{figure}

\begin{figure}[htbp]
    \centering
    \includegraphics[width=0.5\textwidth]{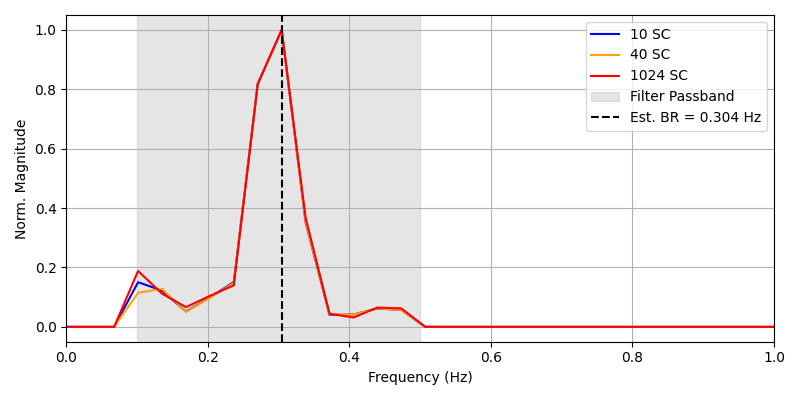}
    \caption{Normalized FFT of a breathing signal for varying subcarrier number, sitting at a desk scenario.}
    \label{fig:compare_breathing}
\end{figure}

\begin{figure}[htbp]
    \centering
    \includegraphics[width=0.5\textwidth]{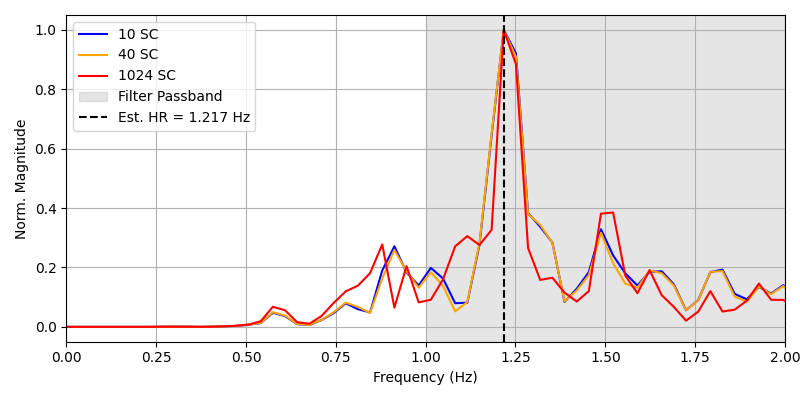}
    \caption{Normalized FFT of a heart signal for varying subcarrier number, sitting at a desk scenario.}
    \label{fig:compare_heart}
\end{figure}

Figure~\ref{fig:compare_phase} shows the unwrapped phase signal over time for three subcarrier configurations. It can be seen that decreasing the number of subcarriers does not negatively influence the quality of the data gathered. Figures~\ref{fig:compare_breathing} and \ref{fig:compare_heart} show the phase signal in the frequency domain, after filtration for the frequencies of breathing and heart beat, respectively. The obtained spectra are highly correlated, especially for the breathing signal. The result of the frequency axis does not change because it does not depend on the bandwidth, but on the time of measurement and pulse interval. 

\begin{figure}[htbp]
    \centering
    \includegraphics[width=0.95\linewidth]{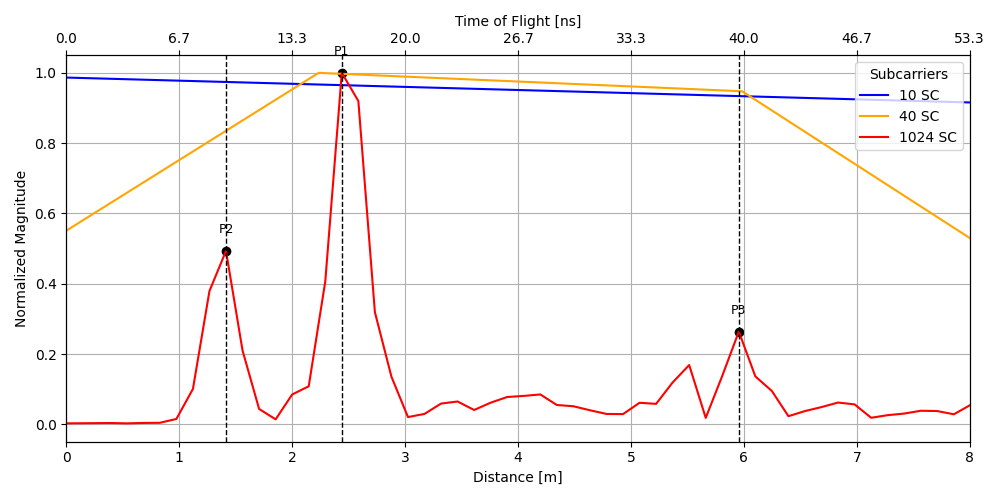}
    \caption{Range profiles for different subcarrier numbers. Three persons were correctly separated only for the 1GHz bandwidth.}
    \label{fig:range_resolution_comparison}
\end{figure}

Figure~\ref{fig:range_resolution_comparison} illustrates how available bandwidth impacts the range resolution. For a scenario with three persons in a single room, only the scenario that utilized the full bandwidth was able to properly separate them all. Therefore, for VS measurement with more than one person inside a room, a higher bandwidth is necessary, as it provides sufficient range resolution. Without differentiating between persons, their VS are present in one signal and are indistinguishable. Even if not all VS are going to be measured, discarding the data coming from other persons is crucial for the measurement accuracy. 

\section{Conclusions and Future Work}

The experiments demonstrated the possibilities and challenges of measuring human vital signs using a mmWave OFDM joint communication and sensing system. Each scenario was designed to measure the system performance under varying conditions. The experiments investigated how the position, orientation, and movement of the person affect the ability to extract heart and breathing-induced body movements accurately. The examined scenarios also investigated how clothing and blankets affect the measurement, whether the radar can measure the vital signs of two persons, and whether the measurement is possible in NLOS conditions. 

The results of the experiments confirm mmWave system’s suitability for contactless monitoring vital signs, particularly breathing, which can be accurately measured in a wide range of scenarios, from a sitting, stationary person to NLOS conditions. Heart rate movement has a lower amplitude than breathing and is more susceptible to noise. Since HR frequencies are typically higher than those of breathing, higher harmonics of breathing can interfere with HR measurement, and the algorithm may incorrectly identify them as HR. As a result, HR measurement is significantly more complex in scenarios involving the movement of the person, NLOS conditions, or orientation that is not in the direction of the radar. The accuracy of the measurement also depends on how deep the person is breathing. In cases of shallow breathing, the results tend to be less precise in more challenging scenarios.

An algorithm utilizing information on the first two or three harmonics of breathing can enhance the measurement of HR by providing a more distinct separation between the harmonics of the breathing signal and HR. It is seen as a potential step forward. Another possible improvement of signal processing lies in tracking the person. Currently, each measurement of VS is taken on a sample lasting from approximately 15 to 40 seconds, during which the person is assumed to remain relatively still. Implementing a person-tracking algorithm could allow dynamic adjustment of the observation point. However, even with such an algorithm, reliably measuring vital signs may be difficult due to overlapping movements. Another potential extension of the proposed signal processing approach could be to use the collected data for training machine learning models to extract VS directly from phase variation signals. Such an algorithm could learn patterns and correlations in the varying phase data and perform better than the regular FFT. Convolutional neural networks are the potential architectures that could be suitable for extracting vital signs from phase variation data due to their ability to analyze sequential dependencies in time series of signals. 

In conclusion, mmWave OFDM radar is one of the most effective technologies for contactless measurement of respiratory rate and HR. VS can be measured through clothing or blankets, making the technology suitable for real-world scenarios. The examined system, after considering privacy aspects, is applicable for medical monitoring in environments where people are mostly stationary, including hospital beds and elderly care facilities, or for sleep monitoring. It can be also used for security applications, such as detecting intrusions or identifying personnel in restricted areas. Furthermore, it can be integrated into a wireless device for simultaneous contactless monitoring of a person and data transmission. We demonstrated that VS can be successfully measured even with narrow channels, corresponding to bandwidths used in consumer-grade Wi-Fi hardware. Under such conditions, the main challenge lies in the system's ability to operate in the presence of multiple people, as the radar cannot distinguish between them without a sufficiently large bandwidth. Another challenge involves VS monitoring in dynamic environments, such as tracking people moving within a room. Future work should also focus on improving algorithm robustness to noise and overlapping movements.

\bibliographystyle{unsrt}
\bibliography{references}

@article{Guan2023,
  author       = {L. Guan and others},
  title        = {Multi-Person Breathing Detection With Switching Antenna Array Based on WiFi Signal},
  journal      = {IEEE Journal of Translational Engineering in Health and Medicine},
  volume       = {11},
  pages        = {23--31},
  year         = {2023},
  doi          = {10.1109/JTEHM.2022.3218638},
  publisher    = {IEEE}
}

@article{Aardal2013,
  author    = {Ø. Aardal and Y. Paichard and S. Brovoll and T. Berger and T. S. Lande and S.-E. Hamran},
  title     = {Physical working principles of medical radar},
  journal   = {IEEE Trans. Biomed. Eng.},
  volume    = {60},
  number    = {4},
  pages     = {1142--1149},
  year      = {2013},
  month     = apr
}

@article{Alizadeh2019,
  author    = {M. Alizadeh and G. Shaker and J. C. M. D. Almeida and P. P. Morita and S. Safavi-Naeini},
  title     = {Remote Monitoring of Human Vital Signs Using mm-Wave FMCW Radar},
  journal   = {IEEE Access},
  volume    = {7},
  pages     = {54958--54968},
  year      = {2019},
  doi       = {10.1109/ACCESS.2019.2912956}
}

@article{engelhardt2024sdr,
  author       = {M. Engelhardt and others},
  title        = {Accelerating Innovation in 6G Research: Real-Time Capable SDR System Architecture for Rapid Prototyping},
  journal      = {IEEE Access},
  volume       = {12},
  pages        = {118718--118732},
  year         = {2024},
  doi          = {10.1109/ACCESS.2024.3447884}
}

@manual{edan-holter,
  title        = {SE-2003/SE-2012 Series Holter System Recorder User Manual},
  author       = {{EDAN Instruments Inc.}},
  year         = {2019},
  url          = {https://solicmedical.com/pdf/82-01.54.456520-1.9%20SE-2003&SE-2012%20Series%20Holter%20System%20Recorder%20User%20Manual-ES.pdf},
  note         = {Accessed: 2025-05-27}
}

@manual{polar-h9,
  title        = {Polar H9 Heart Rate Sensor User Manual},
  author       = {{Polar Electro Oy}},
  year         = {2020},
  url          = {https://support.polar.com/e_manuals/h9-heart-rate-sensor/polar-h9-user-manual-english/manual.pdf},
  note         = {Accessed: 2025-05-27}
}

@article{Shafiq2014,
  author    = {Ghulam Shafiq and Krishna C. Veluvolu},
  title     = {Surface chest motion decomposition for cardiovascular monitoring},
  journal   = {Scientific Reports},
  volume    = {4},
  number    = {1},
  pages     = {1--9},
  year      = {2014},
  month     = {May},
  doi       = {10.1038/srep04402},
  url       = {https://rdcu.be/emxYq}
}

@article{Owda2017,
  author    = {Abdullah Y. Owda and Neil Salmon and Stewart W. Harmer and Sergiy Shylo and Nicholas J. Bowring and Nadia D. Rezgui and Musarat Shah},
  title     = {Millimeter-wave emissivity as a metric for the non-contact diagnosis of human skin conditions},
  journal   = {Bioelectromagnetics},
  volume    = {38},
  number    = {8},
  pages     = {559--569},
  year      = {2017},
  doi       = {10.1002/bem.22083}
}

@inproceedings{Wu2015,
  author    = {Ting Wu and Theodore S. Rappaport and C. M. Collins},
  title     = {The Human Body and Millimeter-Wave Wireless Communication Systems: Interactions and Implications},
  booktitle = {2015 IEEE International Conference on Communications (ICC)},
  year      = {2015},
  month     = {June},
  organization = {IEEE}
}

@ARTICLE{10107610,
  author={Shahbazian, Reza and Trubitsyna, Irina},
  journal={IEEE Access}, 
  title={Human Sensing by Using Radio Frequency Signals: A Survey on Occupancy and Activity Detection}, 
  year={2023},
  volume={11},
  number={},
  pages={40878-40904},
  doi={10.1109/ACCESS.2023.3269843}}

@article{Leenen2024,
  author       = {Leenen, Jeroen P. L. and Ardesch, Vera and Kalkman, C. Johan and Schoonhoven, Lisette and Patijn, Gert A.},
  title        = {Impact of wearable wireless continuous vital sign monitoring in abdominal surgical patients: before-after study},
  journal      = {BJS Open},
  year         = {2024},
  volume       = {8},
  number       = {1},
  pages        = {zrad128},
  month        = jan,
  doi          = {10.1093/bjsopen/zrad128},
  pmid         = {38235573},
  pmcid        = {PMC10794900},
  url          = {https://doi.org/10.1093/bjsopen/zrad128}
}

@article{Weller2018,
  author       = {Weller, Richard S. and Foard, Kelly L. and Harwood, Thomas N.},
  title        = {Evaluation of a wireless, portable, wearable multi-parameter vital signs monitor in hospitalized neurological and neurosurgical patients},
  journal      = {Journal of Clinical Monitoring and Computing},
  year         = {2018},
  volume       = {32},
  number       = {5},
  pages        = {945--951},
  month        = oct,
  doi          = {10.1007/s10877-017-0085-0},
  pmid         = {29214598},
  url          = {https://doi.org/10.1007/s10877-017-0085-0},
  note         = {Epub 2017 Dec 6}
}

@techreport{Clark2023,
  author       = {Clark, M. and Bailey, S.},
  title        = {Single-Use Wearable Wireless Sensors for Vital Sign Monitoring: CADTH Horizon Scan},
  institution  = {Canadian Agency for Drugs and Technologies in Health},
  address      = {Ottawa, ON},
  year         = {2023},
  month        = nov,
  number       = {EN0048},
  note         = {Internet},
  pmid         = {38320082},
  url          = {https://cadth.ca/sites/default/files/pdf/EN0048_SingleUse_Wearable_Wireless_Sensors_Vital_Sign_Monitoring.pdf}
}

@article{Ahmed2015,
  author    = {Ahmed, M. U.},
  title     = {A personalized health-monitoring system for elderly by combining rules and case-based reasoning},
  journal   = {Studies in Health Technology and Informatics},
  year      = {2015},
  volume    = {211},
  pages     = {249--254},
  pmid      = {25980877}
}

@misc{ResearchMarkets2023,
  author       = {{Research and Markets}},
  title        = {United States Vital Signs Monitoring Devices Market Report 2023–2028},
  howpublished = {\url{https://www.researchandmarkets.com/report/united-states-vital-sign-monitor-market}},
  year         = {2023},
  note         = {Accessed: 2025-06-01}
}

@article{alizadeh2019remote,
  author       = {M. Alizadeh and G. Shaker and J. C. M. D. Almeida and P. P. Morita and S. Safavi-Naeini},
  title        = {Remote Monitoring of Human Vital Signs Using mm-Wave FMCW Radar},
  journal      = {IEEE Access},
  volume       = {7},
  pages        = {54958--54968},
  year         = {2019},
  doi          = {10.1109/ACCESS.2019.2912956}
}

@ARTICLE{wifi-posture-sleep,
  author={Liu, Jian and Chen, Yingying and Wang, Yan and Chen, Xu and Cheng, Jerry and Yang, Jie},
  journal={IEEE Internet of Things Journal}, 
  title={Monitoring Vital Signs and Postures During Sleep Using WiFi Signals}, 
  year={2018},
  volume={5},
  number={3},
  pages={2071-2084},
}

@inproceedings{gu2021realtime,
  author    = {Yu Gu and Xiang Zhang and Huan Yan and Zhi Liu and Yusheng Ji},
  title     = {Real-time Vital Signs Monitoring Based on COTS WiFi Devices},
  booktitle = {2021 IEEE International Conference on Bioinformatics and Biomedicine (BIBM)},
  pages     = {1320--1325},
  year      = {2021},
  organization = {IEEE},
}

@article{XUE2023113715,
title = {Accurate multi-target vital signs detection method for FMCW radar},
journal = {Measurement},
volume = {223},
pages = {113715},
year = {2023},
author = {Wei Xue and Rui Wang and Li Liu and Dongchang Wu}
}

@article{lee2019novel,
  title     = {A Novel Vital-Sign Sensing Algorithm for Multiple Subjects Based on 24-GHz FMCW Doppler Radar},
  author    = {Lee, Hyunjae and Kim, Byung-Hyun and Park, Jin-Kwan and Yook, Jong-Gwan},
  journal   = {Remote Sensing},
  volume    = {11},
  number    = {10},
  pages     = {1237},
  year      = {2019},
  publisher = {MDPI},
}

@inproceedings{wang2015novel,
  author       = {S. Wang and T. Liu and H. Zhu and Y. Wang and K. Wu and L. M. Ni},
  title        = {A Novel Ultra-Wideband 80 GHz FMCW Radar System for Contactless Monitoring of Vital Signs},
  booktitle    = {2015 37th Annual International Conference of the IEEE Engineering in Medicine and Biology Society (EMBC)},
  address      = {Milan, Italy},
  pages        = {4978--4981},
  year         = {2015},
  doi          = {10.1109/EMBC.2015.7319509}
}

%\newpage

\begin{comment}
\section{Biography Section}
If you have an EPS/PDF photo (graphicx package needed), extra braces are
 needed around the contents of the optional argument to biography to prevent
 the LaTeX parser from getting confused when it sees the complicated
 $\backslash${\tt{includegraphics}} command within an optional argument. (You can create
 your own custom macro containing the $\backslash${\tt{includegraphics}} command to make things
 simpler here.)
 
\vspace{11pt}

\bf{If you include a photo:}\vspace{-33pt}
\begin{IEEEbiography}[{\includegraphics[width=1in,height=1.25in,clip,keepaspectratio]{fig1}}]{Michael Shell}
Use $\backslash${\tt{begin\{IEEEbiography\}}} and then for the 1st argument use $\backslash${\tt{includegraphics}} to declare and link the author photo.
Use the author name as the 3rd argument followed by the biography text.
\end{IEEEbiography}

\vspace{11pt}

\bf{If you will not include a photo:}\vspace{-33pt}
\begin{IEEEbiographynophoto}{John Doe}
Use $\backslash${\tt{begin\{IEEEbiographynophoto\}}} and the author name as the argument followed by the biography text.
\end{IEEEbiographynophoto}

\vfill
\end{comment}

\end{document}